\documentclass [12pt,aps,a4paper,prb]{revtex4}
\input{epsf}
\usepackage{graphicx}
\usepackage{amssymb}
\usepackage{latexsym}

\begin{document}
\renewcommand{\topfraction}{0.9}
\renewcommand{\textfraction}{0.1}
\renewcommand{\floatpagefraction}{0.75}
\topmargin 0.0cm
\newcommand {\be} {\begin{equation}}
\newcommand {\ee} {\end{equation}}
\newcommand {\bea} {\begin{eqnarray}}
\newcommand {\eea} {\end{eqnarray}}
\newcommand {\D} {\displaystyle}
\newcommand {\fle} { \stackrel{\rightarrow} }
\newcommand {\som} { \stackrel{\wedge} }
\newcommand {\iz} {\left}
\newcommand {\de} {\right}
\newcommand {\etal} { \emph{et al }}
\renewcommand{\theequation} {\arabic{section}.\arabic{equation}}

\setcounter{section}{0}
\begin{center}
{\Large \bf Continuum Theory of Polymer Crystallization}

\bigskip

{\bf Arindam Kundagrami and M. Muthukumar$^{a)}$}

\bigskip
{\it Department of Polymer Science and Engineering \\
University of Massachusetts at Amherst, Amherst, MA 01003}

\begin{abstract}
We present a kinetic model of crystal growth of polymers of finite molecular weight. Experiments help to classify 
polymer crystallization broadly into two kinetic regimes. One is observed in melts or in
high molar mass polymer solutions and is dominated by nucleation control with $G \sim \exp\left( 1/T \Delta T\right)$,
where $G$ is the growth rate and $\Delta T$ is the super-cooling. The other is observed in low molar mass solutions
(as well as for small molecules) and is diffusion controlled with $G \sim \Delta T$, for small $\Delta T$. Our model
unifies these two regimes in a single formalism. The model
accounts for the accumulation of polymer chains near the growth front and invokes an entropic barrier theory to
recover both limits of nucleation and diffusion control. The basic theory applies to both melts and solutions, and
we numerically calculate the growth details of a single crystal in a dilute solution. The effects of molecular weight 
and concentration are also determined considering conventional polymer dynamics. Our theory shows that entropic considerations,
in addition to the traditional energetic arguments, can capture general trends of a vast range of   
phenomenology. Unifying ideas on crystallization from small molecules and from flexible polymer chains emerge from our theory.    
\end{abstract}

\end{center}


\maketitle

\noindent \small a)Author to whom correspondence should be addressed. E-mail:
muthu@polysci.umass.edu

\normalsize 

\section{Introduction}

\setcounter{section}{1}
\setcounter{equation}{0}

There are several different processes involved in crystal growth
from molecules of both low and high molecular weight; they become more complex for flexible macromolecules or polymers. 
Extensive experiments on the growth kinetics of lamellae in solutions and melts helped to classify
the growth rates broadly into two universality classes. In the first, valid for
melt-grown crystals and solution-grown crystals of relatively high molecular weight,
the growth rate $G$ depends exponentially on the variable
$1/T_c \Delta T$\cite{armgol-92,kelped-73,cooman-75,tod-87} as,
\bea
\label{gdtintro}
G \sim \exp \left[ \frac{-\mathcal{P}}{T_c \Delta T}\right],
\eea
where, $\Delta T = T_m - T_c$ is the super-cooling with $T_m$ and  $T_c$, respectively, as the
crystallization temperature and the equilibrium melting temperature,  
and $\mathcal{P}$ is a parameter. In the second class,
valid for solution-grown crystals of relatively low molecular weight, the growth rate
goes linearly with super-cooling as,
\bea
\label{gdtintro1}
G \sim \Delta T,
\eea
for small $\Delta T$. In this paper, we have developed a model that unifies these two apparently
different physical processes and allows us to capture the limiting behaviours of both classes.  

In a highly complex growth phenomenon such as polymer crystallization involving a multitude of processes,
the rate determining factor is the one which impedes the growth more than any other. 
For example, the entanglement effect of interpenetrating polymer chains must be crucial to the kinetics in melt-grown crystals or
solution-grown crystals of high molar mass, whereas free diffusion or transport effects are possibly
dominant in solution-grown crystals of low molar mass.
The generic growth rates in polymer crystallization ($10^{-3}$ to $10 \mu m/hr.$)\cite{kelped-73,cooman-75,tod-87}
are orders of magnitude lower than that expected in a diffusion-limited crystal growth 
($10$ to $10^5 \mu m/hr.$)\cite{lan-80,dougol-88,maletl-95,lackos-99,deuler-88}.
This immeditately suggests the presence of some sort of a barrier or 'entropic tension' at or near the growth front of
a polymer crystal. In quantitative terms,
Eqn. (\ref{gdtintro}) strongly indicates of an underlying nucleation
process that has been addressed by many 
theories\cite{LH-76,sandim-71,armgol-92,fratos-61}, and primarily by Lauritzen and Hoffman. 
A typical barrier height for a two-dimensional nucleation
(as for polymer crystallization, in which crystals do not grow in the fold-direction) goes as $1 / \Delta F$, where
$\Delta F$ is the free energy change per unit volume of phase change.
That implies a nucleation rate of form Eqn. (\ref{gdtintro}) (assuming a less
debated proportionality between $\Delta F$ and $\Delta T$).
The Lauritzen-Hoffman theory
(the LH theory) generalizes this surface nucleation concept to incorporate chain folding by proposing a distribution of crystal thickness
with a cut-off minimum.
The same distribution is integrated over to calculate 
the average thickness as an observable and to achieve the temperature dependence as in Eqn. (\ref{gdtintro}). 
Sanchez and DiMarzio (the SD theory)\cite{sandim-71}   
further considered the role of cilia (dangling chain ends) in the nucleation process,
but their analysis is broadly in line with the LH theory. 

To put things in perspective, it can be recalled that there are no significant barriers during the growth stage for small molecules, and 
the generic growth rate after initial nucleation can be expressed, based on detailed balance arguments, as:
\bea
\label{smallmolintro}
G_{small} \sim g \left[ 1 - \exp \left( \frac{- v \Delta H \; \Delta T}{k_B T_c}\right) \right],
\eea 
where, $\Delta H$ and  $v$ are, respectively, the enthalpy per unit volume and the minimum volume element of crystallization,  
and $g$ is very weakly dependent on temperature. $v$ is also involved
in the factor ${\mathcal P}$ in Eqn. (\ref{gdtintro}) heavily affecting the growth rate.
For small under-coolings
($\Delta T \rightarrow 0$) the growth rate obeys $G \sim \Delta T$ (Eqn. (\ref{gdtintro1}) ), the linear relationship widely
observed\cite{giljac-77} in small molecule crystallization and known to indicate the thermal 'roughness' regime. 
This behaviour of low molecular weight materials prompted Sadler and Gilmer (the SG theory) to suggest 
that\cite{sad-83,sadgil-84,sad-87} polymer crystal growth is 
driven by kinetic roughening rather than nucleation. The SG theory
assumed that the smallest attaching units can be fractional stems, and roughness  
is inevitable if the surface free energy densities are of order $k_B T_c$. This theory
conceived of a barrier resulting from the interruption  of growth due to pinning and subsequent removal of 
short stems, which are constrained by the connectivity of a longer chain.
The roughness theory was hard to verify 
due to a lack of experimental evidence for roughening transitions\cite{armgol-92} as most polymer
crystals are observed facetted. 
Moreover, comparison 
shows that the growth rate
changes by only one order of magnitude in roughness dominated growth (SG) as opposed to
three in nucleation dominated growth (LH) for a typical range of super-cooling. 

In addition to temperature, the other two major variables that affect polymer crystal growth significantly are 
polymer concentration $C$
and polymer molecular weight $M_w$. 
The many diverse ways by which concentration and molecular weight
influence the growth of polymer crystals can be summarized in the
following relationship:
\bea
\label{gcmintro}
G \sim C^\gamma M_w^{-\mu} f(T_c),
\eea
where, a low ($\ll 1$), constant value of $\gamma$ observed for high molar mass polymers
lends support to barrier control at the growth front.
For solution grown crystals of very low molecular weight,
$\gamma$ is much higher - sometimes close to or even larger than 1\cite{kelped-73,jiamut-06}, suggesting uninterrupted diffusion-limited growth.
The effective exponent $\mu$, however, is not constant even for very low
concentrations. It can assume positive or negative
values depending on the range of $M_w$ investigated\cite{cooman-75} and is generally a complicated function of experimental
variables\cite{okui-02,okui-03,okui-05}.

\begin{figure}[ht] \centering
\hspace*{1cm}{\epsfxsize= 14cm \epsfbox{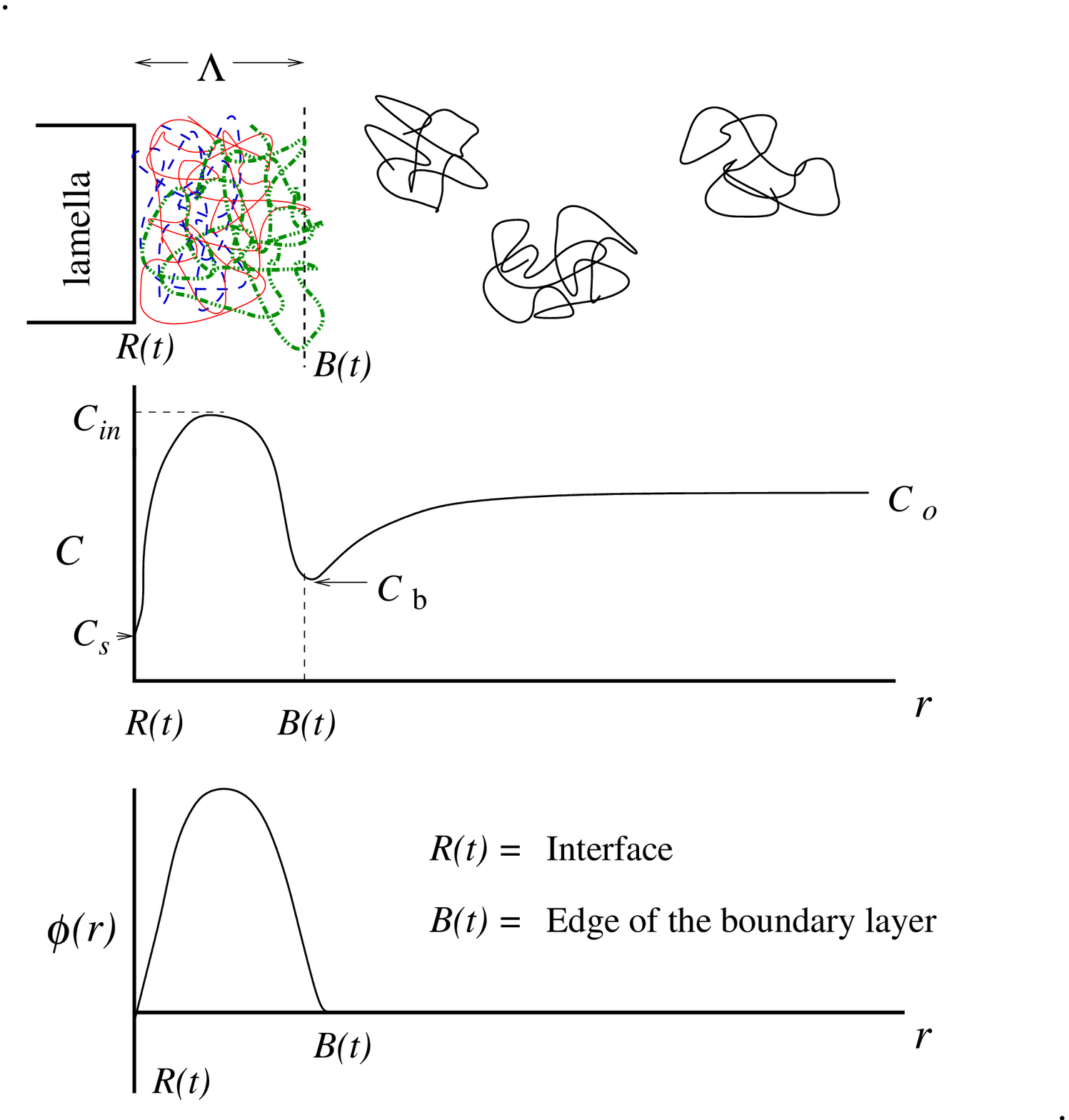}}
\bigskip
\vspace*{-0.5cm}
\vspace*{2cm}
\caption{Kundagrami \it{et al.}, JCP}
\label{paper1-fig1}
\end{figure}

Additional insight has been gained from an impressive wealth of microscopic study of the early stages
of chain folding and crystal growth through Langevin dynamics 
simulations\cite{kavsun-93,mut-96,liumut-98,fujsat-98,mut-00,mutwel-00,welmut-01,yam-01,dukmut-03,mut-03,mut-04,yam-04} as well as
Monte Carlo techniques\cite{doyfre-99,doy-00,somrei-00}.
An overwhelming majority of simulations show accumulation
of  multiple polymer chains near the growth front even for very dilute solutions, especially for longer chains. 
This temporal congestion of molecules invokes a strong possibility that these
molecules would be subject to an entropic pressure  and excluded volume effect that can significantly 
control the growth kinetics. 
Polymers of low molecular weight in solution\cite{jiamut-06,mut-07} as well as relatively larger rigid small 
molecules such as $GeO_2$ or $P_2O_5$\cite{wunspl-80}
follow Eqn. (\ref{smallmolintro}) suggesting that the {\it flexibility} of large polymer chains, not simple energy considerations, might be
key to the ordering behaviour in Eqn. (\ref{gdtintro}). Finally, many of these 
simulations\cite{doyfre-99,doy-00} and experiments do not support much variation 
of crystal thickness for a fixed supercooling, questioning the LH argument of a distribution and minimum of the same.   
Having considered these,
we have developed a continuum coarse-grained theory of polymer crytallization, with a focus on the moleculer 
details near the growth front for relatively larger chains. Mainly based on the
concept of an entropic pressure, the model unifies nucleation and diffusion control behaviours as in, respectively, 
Eqns. (\ref{gdtintro}) and (\ref{smallmolintro}).
The key aspects of the theory are as follows. 
If we consider a single polymer crystal with a specific thickness, we assume that there would be 
accumulation of chains near the growth front resulting from an apparent nucleation control. As a result of this accumulation,
which happens regardless of the bulk concentration,
the local monomer concentration increases considerably, to a value ($C_{in}$) much higher than the rest of the system ($C_0$). 
This happens in a narrow layer region adjacent to the front 
(called the 'boundary layer' region henceforth)[Fig. \ref{paper1-fig1}]. 
Due to their higher concentration, these interpenetrating unadsorbed polymer chains have reduced number of configurations available that
creates an entropic barrier within the boundary layer. 
The diffusive
macromolecules must negotiate this entropic barrier before their attachment to the crystal face. The
boundary values of polymer concentration $C_s$ and $C_b$, respectively, at the interface ($R(t)$) and
the outer boundary layer edge ($B(t)$), can be different depending on the nature of the barrier. 
Considering an appropriate free energy associated with the entropic barrier, 
and assuming the barrier layer thin enough to let
the radial fluxes at the interface and the edge of the layer be equal, the following growth rate is
predicted:
\bea
\label{growthintro}
G \sim C_0 D_{in} \exp \left( \frac{-\mathcal{P}}{T_c \Delta T} \right) \left[ 1 - \exp \left( -{\mathcal Q} \Delta T \right) \right],
\eea
where, $D_{in}$ is the diffusivity of the polymers inside the barrier layer.
$\mathcal{P}$ and $\mathcal{Q}$ are system variables very weakly dependent on temperature, concentration or molecular weight.
Notable points on the above equation include:  
a) the $1/\Delta T$ factor is recovered in this theory from the proposed entropic barrier, 
b) the limiting behaviours of both nucleation and diffusion control are obtained in a single expression and
c) the dense boundary layer enforces a dynamics different to that of the bulk system
which affects the concentration and molecular weight behaviours significantly.

As an illustration of the barrier theory we numerically calculate 
the growth of a single crystal in dilute solution. In the numerical model, we conceive of a cylindrical lamella with  circular
cross-section and fixed thickness $L$[Fig. \ref{paper1-fig2}]. The fold-area 
dimension is typically much larger than the thickness in course of growth 
giving the crystal a shape like a tablet.  
The single crystal is embedded in a bath of diffusing polymer molecules, and
during the growth process a new chain can
only attach to the lateral surface, but not to the top and bottom surfaces (the fold surfaces) of the lamella. The 
crystal-solution interface and the edge of the boundary layer
are calculated as functions of time, considering at $B(t)$ a boundary condition involving a rate-constant, which
directly depends on the free energy function inside the boundary layer. The values of the effective exponents
$\gamma$ and $\mu$ follow from the dependencies of $D_{in}$ and $T_m$ on concentration and molecular weight.

The rest of the paper is organized as follows: the theory is detailed in Sec. II; the analytical and numerical results are 
presented in Sec. III; conclusions are summarized in Sec. IV.

\section*{II. THEORY}
\subsection*{A. The continuity equation:}
\setcounter{section}{2}
\setcounter{equation}{0}

The theoretical model considers the growth of a lamella of a prescribed thickness $L$ and
the shape of a cylindrical tablet with circular cross-section (Fig. \ref{paper1-fig2}). 
The lamella grows radially outward in a medium containing polymer chains at an initial uniform concentration $C_0$.
The analysis of the simplest scenario of diffusion limited growth starts with the time-dependent diffusion equation 
in cylindrical polar coordinates with the concentration ($C$) of the
monomers as the diffusion variable.  
The azimuthal symmetry inherent in the system allows us to write,
\begin{eqnarray}
\label{diffeq}
D_{out} \left(\frac{\partial^2 C}{\partial r^2} + \frac{1}{r} \frac{\partial C}{\partial r} +
\frac{\partial^2 C}{\partial z^2} \right) = \frac{\partial C}{\partial t},
\end{eqnarray}
where, $D_{out}$ is the diffusion constant and $C$ is the concentration of the material in the outer region 
(all $z$ for $r > B(t)$ and $|z| > L/2$ for $r < B(t)$).
We invoke the mass balance equation at the interface ($r = R(t)$) as, 
\begin{eqnarray}
\label{massbal}
(C_{solid} - C_s) \frac{dR}{dt} = \mbox{flux at the interface},
\end{eqnarray}
where, the concentration of monomers is $C_{solid}$ in the crystalline (solid) phase and $C_s$ in the uncrystallized (solution or melt) phase at the
interface.  $R(t)$ is the radius of the lamellar crystal as well as the location of the crystal-solution(melt) interface
at time $t$ and $B(t)$ is the location of the outer edge of the boundary layer.
The growth or, in quantitative terms, the radius $R(t)$ as a function of time is calculated from Eqns. (\ref{diffeq}) and (\ref{massbal})
which are solved in accordance with the following boundary conditions: a)
\bea
\label{bc1}  
\frac{\partial C}{\partial n} = 0 
\eea
at the boundaries of the system (the container in which the crystal is growing), where, $n$ is the direction normal to the surface; 
b) 
\bea
\label{bc2}
\frac{\partial C}{\partial z} = 0
\eea
at the fold surfaces (the top and bottom surfaces of the cylindrical tablet) given by $|z| = L/2$ for $0 < r < R(t)$; 
and c)
\bea
\label{bc3}
C=C_s(z)
\eea
at the interface
layer given by $r = R(t)$ for $0 < |z| < L/2$. 
The reflecting boundary condition (boundary condition b)) is adopted to model 
the physical situation in which the polymer molecules are denied attachment to the fold surfaces of the lamella.
Unlike many other theories that deal with a particle-wall type interactions\cite{myelyt-99},
we do not assume a perfect sink boundary condition in which $C = 0$ at the wall.
In our model, $C = C_s \neq 0$ at the interface at $r = R(t)$.  For
crystallization at finite temperatures, desorption and the preservation of detailed
balance at the interface preclude the use of a perfect sink (or an immobilizing) condition.
Therefore, in general, the concentration of the mobile molecules at the interface ($C_s$) is an unknown variable
in our theory.

To replace the boundary condition at the interface at $R(t)$ with a boundary condition at the edge of the
boundary layer at $B(t)$, we start with the general expressions for the current term in diffusion equation. 
The generic continuity equation will be
\bea
\label{conteq}
\frac{\partial C}{\partial t} = \bf{\nabla} \cdot \bf{J},
\eea
where the flux $\bf{J}$ is of the form
\bea
\label{flux}
J_i = C V_i - D_{ij} \partial_j C + \frac{C}{k_B T} D_{ij} F_j,
\eea
where, $i,j$ are indices for the components $r,\phi,z$.
The first term describes convection as a function of the concentration $C$ and mass velocity
$V$. The second term represents the driving force due to the concentration gradient, 
where $D_{ij}$ are the diffusivity tensors. The third term describes an external force that
can be conveniently represented by a potential or a free energy.

For the slow process of polymer crystallization, the
convective current is generally negligible. Azimuthal symmetry ensures that $J_\phi$
must be zero. Cross diffusion is negligibly small rendering $D_{rz} = D_{zr} =0$. In conjunction with
these criteria, the standard expressions for the gradient and the divergence in cylindrical polar
coordinates in Eqn. (\ref{flux}) yield the following current terms:
\bea
\label{fluxrz}
J_r &=& - D_{rr} \frac{\partial C}{\partial r}  + \frac{C}{k_B T} D_{rr} F_r \nonumber \\
J_\phi &=& 0 \nonumber \\
J_z &=& - D_{zz} \frac{\partial C}{\partial z} + \frac{C}{k_B T} D_{zz} F_z,
\eea
where, the diffusivity tensor $D_{ij}$ and the external force $F_j$ are written in their component forms.
We would be interested in steady-state growth only rendering $\frac{\partial C}{\partial t} = 0$.
Combining Eqns. (\ref{conteq}),(\ref{fluxrz}) and the steady-state condition, we get
\begin{widetext}
\bea
\label{conteqstead}
{\bf \nabla} \cdot {\bf J} &=& \frac{1}{r} \frac{\partial}{\partial r} \left( r \left\{ - D_{rr} \frac{\partial C}{\partial r}
+ \frac{C}{k_B T} D_{rr} F_r \right\} \right) + \frac{\partial}{\partial z} \left( - D_{zz} \frac{\partial C}{\partial z}
+ \frac{C}{k_B T} D_{zz} F_z\right) \nonumber \\
&=& 0.
\eea
\end{widetext}

The analysis detailed above is a very general description of a crystallization process governed by  reaction-diffusion
equations (Eqn. (\ref{diffeq}) alongwith Eqn. (\ref{massbal})). Eqn. (\ref{conteqstead}),
although applicable to a variety of cases regardless of the size and structure of the molecules, is more appropriate
for slowly diffusing linear homopolymers as mentioned above. The most challenging aspect of this scheme
is to determine the concentration $C_s$ at the interface, more so when large molecules and possible
entanglements result in complex dynamics near the growth front. To deal with it in our model,
we propose a method considering a dense boundary layer at the growth front, in which the polymer molecules 
are subject to an entropic barrier and the polymer dynamics is different than in the 
rest of the system. Before proceeding further with our analysis, we now provide an outline 
of our boundary layer formalism.

\subsection*{B. The boundary layer:}

\begin{figure}[ht] \centering
\hspace*{1cm}{\epsfxsize= 14cm \epsfbox{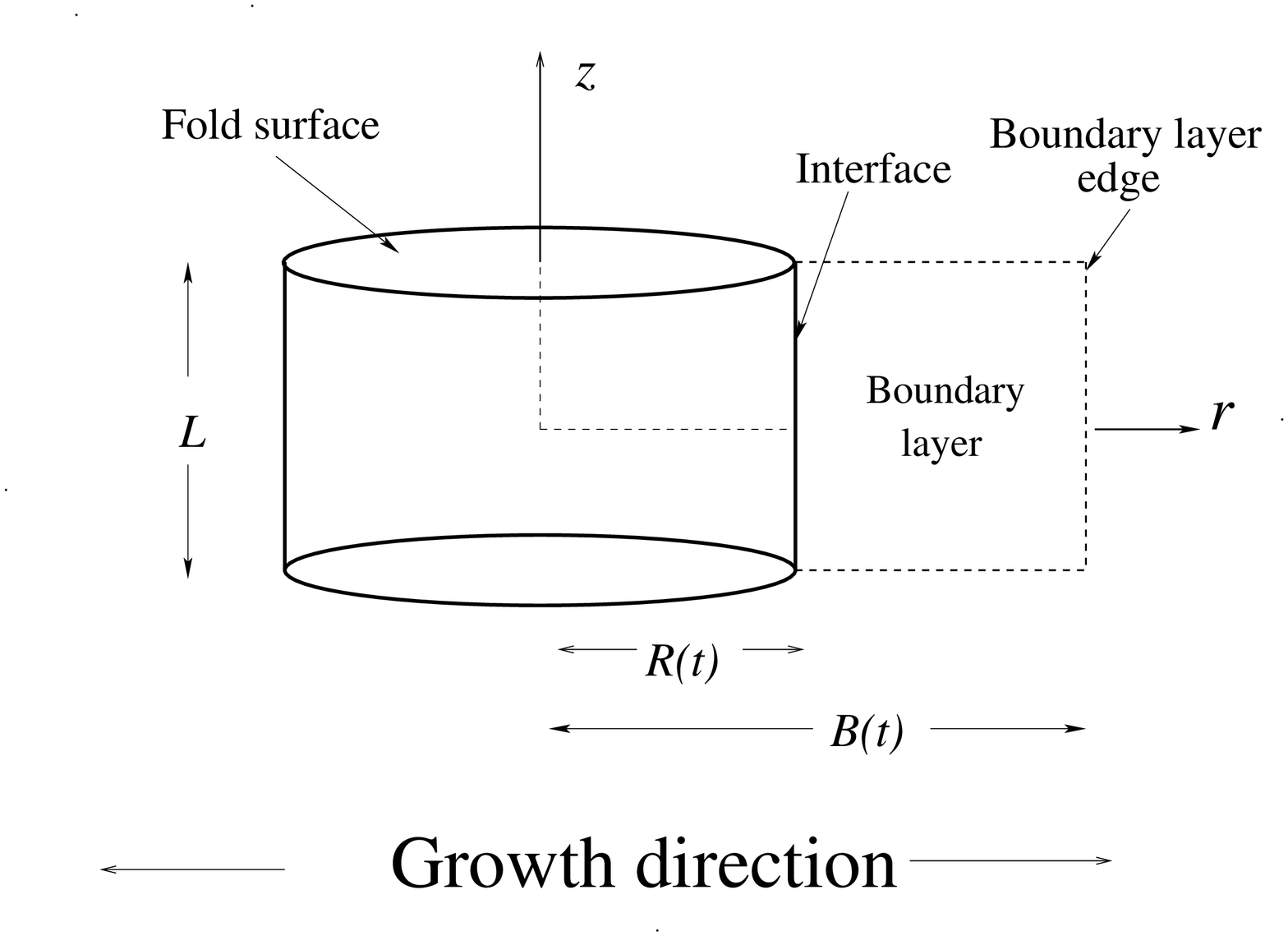}}
\bigskip
\vspace*{-0.5cm}
\vspace*{2cm}
\caption{Kundagrami \it{et al.}, JCP}
\label{paper1-fig2}
\end{figure}

Schematically, we specify two different regions 
in the uncrystallized part of the system - namely, the 'outer region' 
(all $z$ for $r > B(t)$ and $|z| > L/2$ for $r < B(t)$, Fig. \ref{paper1-fig2})
with the bulk polymer concentration 
and the 'boundary layer region' ($|z| < L/2$ for $R(t)<r<B(t)$) with concentration much higher than the bulk value.  
The polymer molecules are subject to free diffusion only in the 'outer region', whereas they experience entropic force inside the
'boundary layer'. These two apparently different processes are reconciled by matching the boundary conditions at 
the common 'interface' of these two regions($r = B(t)$). Treatment of the boundary layer must include the effect of the
free energy barrier  resulting from the entropic pressure adjacent to the growth front.

Before further simplifying the expression in Eqn. (\ref{conteqstead}), we elaborate on  the assumptions made on polymer flow
inside the boundary layer. If we focus on the attachment mechanism of a single polymer chain, 
the rectangular area at the growth front in Fig. \ref{paper1-fig2}
can be treated as a 'hot-seat'. Diffusing polymer molecules, while trying
to attach to the growth front (or the interface at $R(t)$), would have to occupy the 'hot-seat'
prior to attachment. The polymer molecules in this 'hot-seat' are subject to the 
entropic pressure, and therefore, our primary assumtion would be 
the barrier force ${\bf F}$ in Eqn. (\ref{conteqstead}) is non-zero inside and negligible outside 
this 'hot-seat' region. 
In general, however, ${\bf F}$ will have non-zero components $F_r$ and $F_z$. We notice that, within the 'hot-seat',
some of the molecules would already resemble the structural morphology of a full grown crystal
and therefore the corresponding stems will mostly be parallel to the growth front (Fig. \ref{paper1-fig3}a). 
Any diffusing molecule trying to attach from the $z$-direction, regardless of 
the orientation of its stems, will have minimal penetration within
the layer (similar to the fold surfaces). Therefore, we can safely ignore
$F_z$ inside this boundary layer. Considering this and 
the flux in the $z$-direction at the mid-layer of the lamella($z = 0$) which follows 
\bea
\label{bcmidlayer}
\frac{\partial C}{\partial z} \left( r, 0 \right) = 0, 
\eea
allowed by symmetry,
we can reasonably argue that the flux in the $z$-direction
with respect to that in  $r$-direction can be ignored within this boundary layer region.

Instead of solving the generalized diffusion equation 
(Eqns. (\ref{conteq}) with (\ref{flux}) ) inside the boundary layer in which
the barrier force ${\bf F}$ is active, we propose to set up a boundary condition at the outer layer edge($r = B(t)$)
as a function of relevant physical variables.
Now that we have argued that the diffusion as well as the entropic force in the $z$-direction 
are negligible with respect to their $r$ counterparts within the boundary layer,
the third and fourth terms in Eqn. (\ref{conteqstead}) may be ignored, and
the steady-state continuity equation inside the layer simplifies to
\bea
\label{eqsimple}
 \frac{1}{r} \frac{\partial}{\partial r} \left( r \left\{ - D_{rr} \frac{\partial C}{\partial r}  
+ \frac{C}{k_B T} D_{rr} F_r \right\} \right) = 0.
\eea
As a much higher concentration inside the layer is anticipated, we set
the diffusion coefficient $D_{rr} = D_{in}$, which is different from the bulk diffusivity, $D_{out}$.
Integrating the above equation in $r$ once, we obtain
\bea
\label{eqint1}
 r \left\{ - D_{in} \frac{\partial C}{\partial r}  
+ \frac{C}{k_B T} D_{in} F_r \right\} = {\mathcal C}(z)
\eea
for $r \neq 0$. Identifying the quantity in the parenthesis as the radial
flux of polymer chains $J_r (r)$, we determine the integration constant ${\mathcal C}(z)$
and express Eqn. (\ref{eqint1}) as
\bea
\label{eqint1const}
 - D_{in} r \frac{\partial C}{\partial r}  
+ \frac{C r}{k_B T} D_{in} F_r  = J_r\big|_{B(t)} B(t),
\eea
where, $r = B(t)$ is the edge of the boundary layer. Treating the diffusion
coefficient as a general $r$-dependent quantity $D_{in} (r)$, multiplying by
the integrating factor of form 
$$\exp \left[ \int_{R(t)}^{B(t)} \left[ - \left( \frac{F_r (r)}{k_B T} \right) dr \right] \right]$$ and integrating once more
over $r$, we get
\begin{widetext}
\bea
\label{eqint2}
C(r) - C_s = - J_r\big|_{B(t)} B(t) \exp \left( - \frac{\phi (r)}{k_B T} \right)
\int_{R(t)}^{B(t)} \frac{\exp \left( \frac{\phi (r')}{k_B T} \right)}{D_{in}(r') r'}
dr'.
\eea 
\end{widetext}
Here, we have enforced the boundary condition in Eqn. (\ref{bc3}). In addition, we have introduced a potential
$\phi (r)$ corresponding to the force ${\bf F} (r)$ given by the equation
\bea
\label{phipotential}
F_r (r') = - \frac{\partial \phi (r')}{\partial r'}.
\eea

Outside the boundary layer region ${\bf F}$ is zero. Polymer molecules
undergo free diffusion in this region and the reduced equation (Eqn. (\ref{conteqstead})) becomes,
\bea
\label{conteqout}
\frac{1}{r} \frac{\partial}{\partial r} \left( - D_{rr} r \frac{\partial C}{\partial r}  
\right) + \frac{\partial}{\partial z} \left( - D_{zz} \frac{\partial C}{\partial z} 
\right) = 0. 
\eea
Comparing with the steady state equation $\partial_i J_i = 0$, we identify the two
terms in the parentheses above as fluxes in the $r$ and $z$ directions, respectively. Therefore,
the radial flux in the outer region can be written as,
\bea
\label{outerflux}
J_r = &&- D_{out} \frac{\partial C}{\partial r} \:\:\:\mbox{for} \nonumber \\ 
                                                     &&r \geq B(t)\;\;\mbox{and} \;\; r < B(t); |z| < L/2.
\eea
At this point we enforce the continuity condition at the edge of the boundary layer ($r = B(t); |z| < L/2$) 
for both radial flux $J_r$ and concentration $C$ assuming they are equal, respectively,  at both sides of the boundary. 
Evaluating expressions valid for inside and outside of
the boundary layer from Eqns. (\ref{eqint2}) and (\ref{outerflux}), respectively, 
and comparing them at the edge of the layer ($r = B(t)$), we obtain
\begin{widetext}
\bea
\label{equateinout}
- D_{out} \frac{\partial C}{\partial r}\bigg|_{B(t)}
= - \Big[ C(B(t)) - C_s \Big] \exp \left( \frac{\phi (B(t))}{k_B T} \right)
\bigg/ B(t)\int_{R(t)}^{B(t)} \frac{\exp \left( \frac{\phi (r')}{k_B T} \right)}{D_{in}(r') r'} dr'.
\eea
\end{widetext}
Rearrangement of terms yields
\bea
\label{blbc}
\frac{1}{C(B(t)) - C_s}\frac{\partial C}{\partial r}\bigg|_{B(t)} = K, 
\eea
where,
\bea
\label{rateconstant}
K = \exp \left( \frac{\phi (B(t))}{k_B T} \right)
\bigg/ D_{out} B(t)\int_{R(t)}^{B(t)} 
\frac{\exp \left( \frac{\phi (r')}{k_B T} \right)}{D_{in}(r') r'} dr'
\eea
is conceived as the effective rate-constant for this reaction-diffusion mechanism. There are several features of the expression
of $K$ worthy of note. As suggested at the beginning of the sub-section, we have evaluated $K$ as a function
of the variables inside the boundary layer, although it can be used as a measure of the flux (Eqn. (\ref{blbc})) just
at the edge of the layer. In the process of deriving $K$, we have eliminated all complexities of solving the full reaction-diffusion
equation in the layer region ($R(t) < r < B(t); |z| < L/2$). Moreover, we have gained substantial insight into the problem 
just by producing a boundary condition for the bulk solution (the 'outer region') in terms of the variables inside the layer region.  

\subsection*{C. The entropic barrier}

\begin{figure}[ht] \centering
\begin{minipage}{15cm}
\vspace*{1.2cm}
\hspace*{0.0cm}\textsf{\textbf{(a)}}\\
\vspace*{-1.0cm}
\hspace*{1.5cm}{\epsfxsize= 15cm \epsfbox{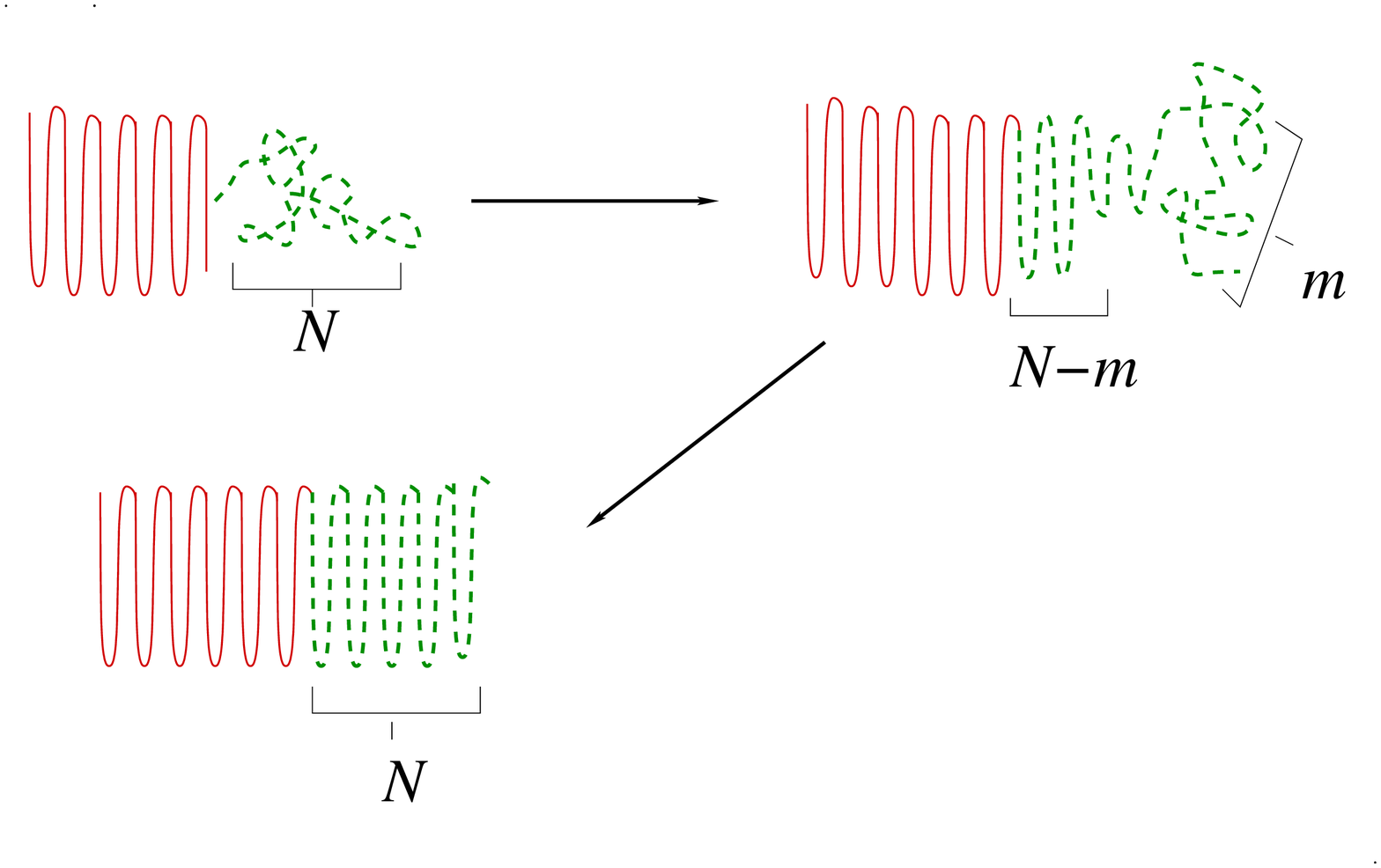}}
\vspace*{1.2cm}
\hspace*{0.0cm}\textsf{\textbf{(b)}}\\
\vspace*{-1.0cm}
\hspace*{1.5cm}{\epsfxsize= 10cm \epsfbox{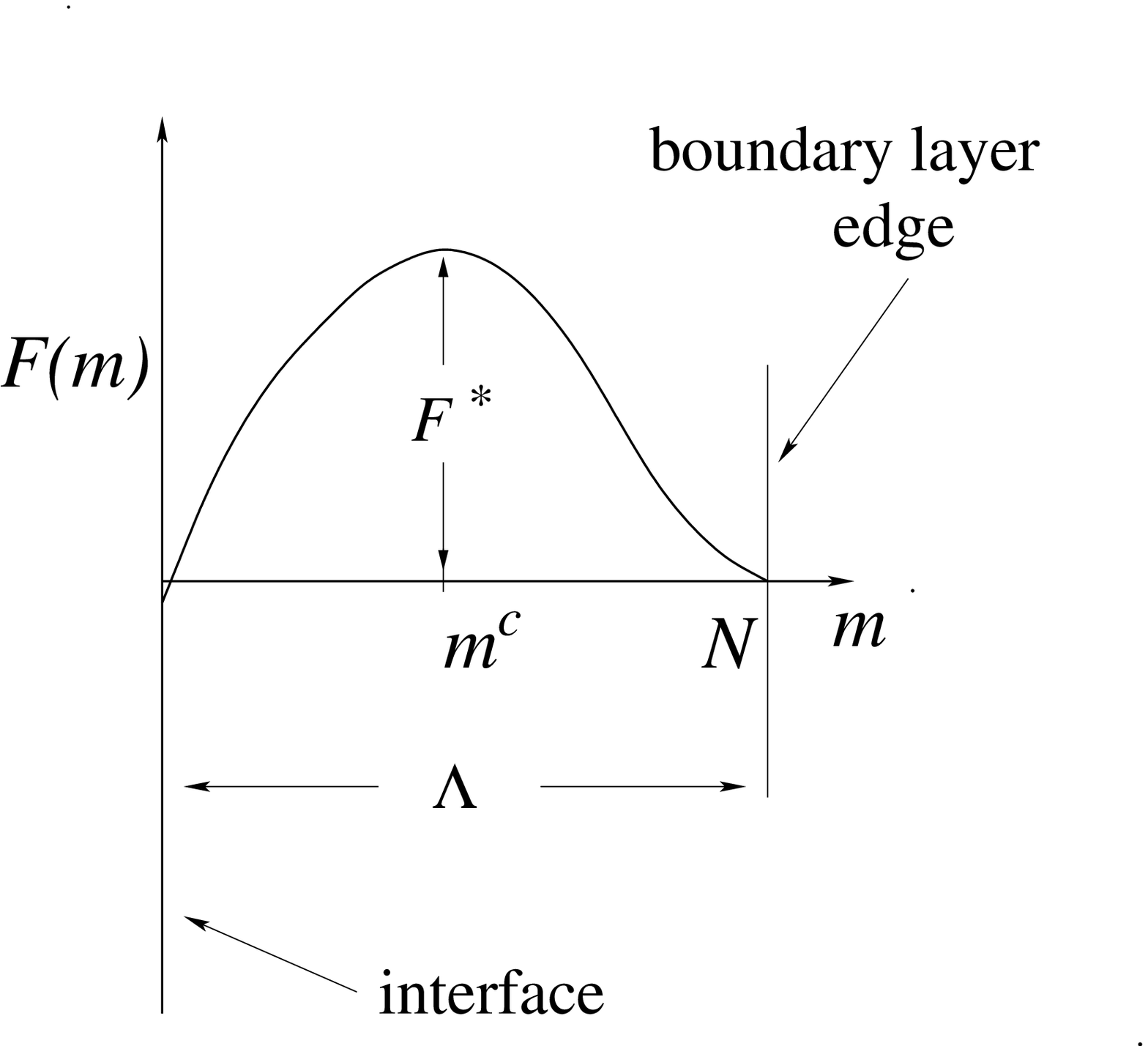}}
\caption{Kundagrami \it{et al.}, JCP}
\label{paper1-fig3}
\end{minipage}
\end{figure}

\begin{figure}[ht] \centering
\hspace*{1cm}{\epsfxsize= 14cm \epsfbox{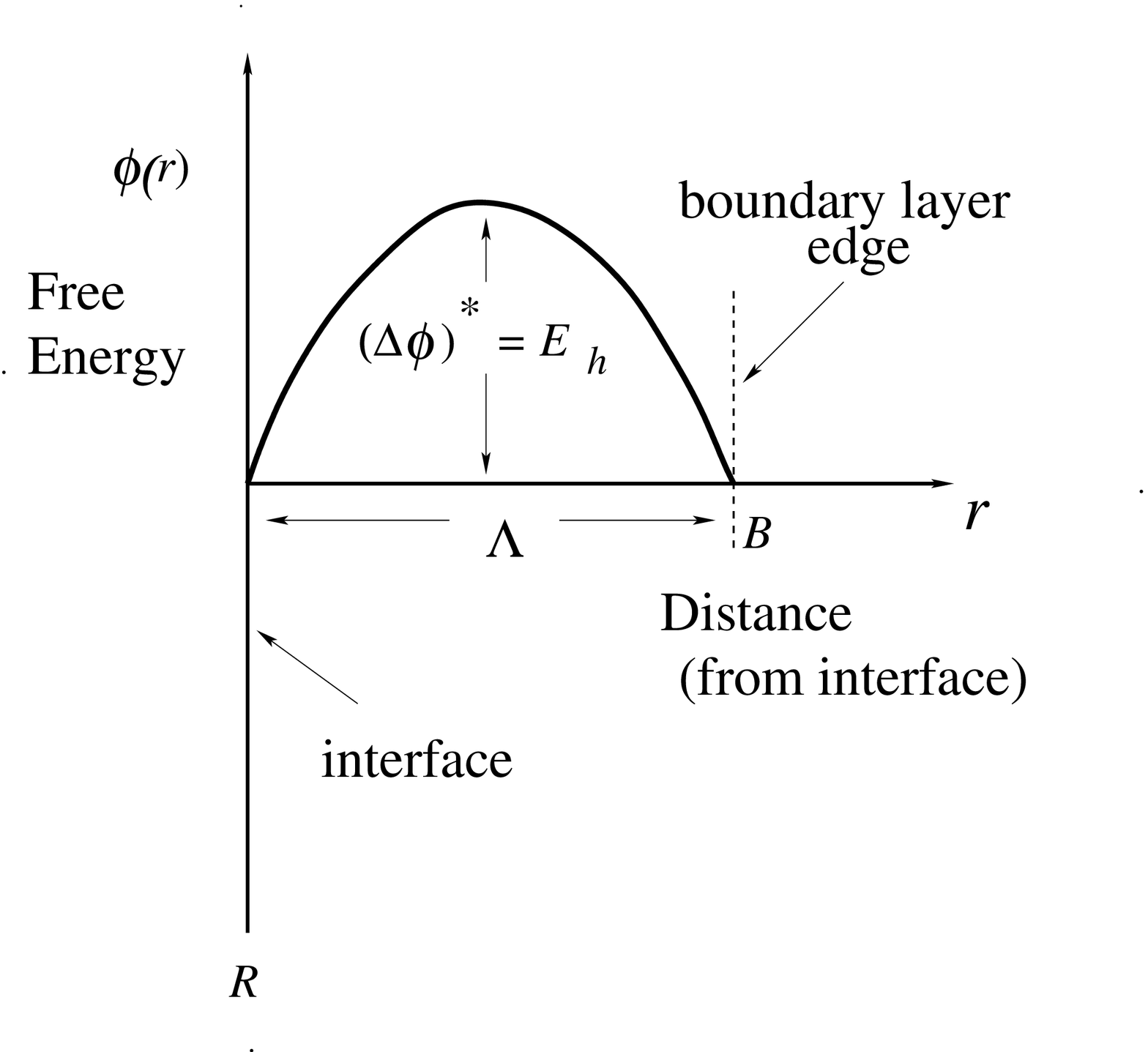}}
\bigskip
\vspace*{-0.5cm}
\vspace*{2cm}
\caption{Kundagrami \it{et al.}, JCP}
\label{paper1-fig4}
\end{figure}

The novel concept of a barrier created due to the accumulation of polymer molecules
at the growth front is the most notable aspect in our theory. Fig. \ref{paper1-fig1} 
illustrates the key features of the barrier.
To determine this barrier quantitatively in terms of the potential $\phi (r)$, we
model it with a free energy functional. 
The number of monomers incoporated
into the solid is assumed to be roughly proportional to the distance up to which the molecule has penetrated the barrier.
The polymer molecule that diffuses through the bulk, negotiates the barrier
and tries to get registered in the crystal (Fig. \ref{paper1-fig3})
starts to feel the barrier force
when at least one monomer enters the boundary layer region. It will cease to
feel the force once the whole of it is incorporated into the solid. The barrier
force will be maximum when roughly half of the molecule is in the solid and the
other half is still in the layer region. Therefore,
Gaussian or parabolic profiles (Fig. \ref{paper1-fig4}) for the layer free energy 
might be natural choices. Choosing a parabola makes the calculation easier, although
the saddle-point method (described with analytical results in Section.III) illustrates that the choice 
has little effect on key results. The expression of the parabola
in Fig. \ref{paper1-fig4} would be given by
\bea
\label{parabola}
\phi (r) = E_h - \frac{E_h}{\Big[ (B(t) - R(t))/2 \Big]^2} \left( r - \frac{B(t)}{2} \right)^2,
\eea
where, $E_h$ is the maximum height of the parabola and $\phi (r)$ is the free energy function.
$E_h$ will be identified as the barrier height henceforth. 

To determine the barrier height we write the free energy in terms of the number of monomers already
incorporated in the crystal (Fig. \ref{paper1-fig3}b) as, 
\bea
\label{energym}
F_m = - (N - m) \Delta F + \sigma_g \sqrt{N -m} + (1 - \gamma') \ln m,
\eea
where, $N$ is the total number of monomers in the molecule, $m$ is the
number of monomers still unattached to the solid front, $\sigma_g$ is
a general 'surface free-energy' quantity, $\Delta F$ is the
bulk energy gain per unit volume of crystallization  and $\gamma'$ is the surface critical exponent\cite{muttrans-99}.
Maximization of the free energy in terms of $m$ yields,
\bea
\label{barrierheight}
F_m^\star \sim \frac{1}{(\Delta F )}.
\eea 
We identify $F_m^\star$ with $E_h$, the barrier height, in Eqn. (\ref{parabola}) and obtain the final
expression for the barrier potential,
\bea
\label{barrierpotential}
\phi (r) = \frac{A}{\Delta F} - \frac{A / \Delta F}{\Big[ (B(t) - R(t))/2 \Big]^2} \left( r - \frac{B(t)}{2} \right)^2,
\eea 
where, $A$ is a quantity dependent on the system variables but not too sensitively dependent on temperature, concentration
and molecular weight. $A$ can be treated as a parameter in the above equation for basic growth studies that are compliant with
typical experiments.

Before proceeding with our analysis, we summarize the theoretical scheme detailed above. We aimed to solve 
the time-dependent continuity equation
(diffusion equation - Eqn. (\ref{diffeq})) in the bulk polymer solution subject to the boundary conditions as specified
at the wall of the container (Eqn. (\ref{bc1})), the fold surfaces of the lamella (Eqn. (\ref{bc2})) and the
edge of the proposed boundary layer (Eqn. (\ref{blbc})). The monomer concentration has been treated as the diffusion variable,
and the rate-constant $K$ at the boundary layer edge  is derived in Eqn. (\ref{rateconstant}), 
in which the free energy functional $\phi$ is given by Eqn. (\ref{barrierpotential}). The 
growth of the interface has to be calculated from the mass-balance equation (Eqn. (\ref{massbal})) 
assuming that fluxes are equal at the interface and at the outer edge of the boundary layer.

\section*{III. RESULTS AND DISCUSSION}
\setcounter{section}{3}
\setcounter{equation}{0}

\subsection*{A. Analytical results:}

For the above described model, it is fairly straightforward to derive 
the basic growth laws.  The growth rate $\frac{dR}{dt}$ of the interface at $R(t)$ is given by the
mass-balance equation (Eqn. (\ref{massbal})) in which the flux is calculated at the interface. On the
other hand, the flux at the edge of the boundary layer at $B(t)$ is given by 
the boundary condition in Eqn. (\ref{blbc}). Combining the two in conjunction with the major assumption above,
we obtain the expression 
for the growth rate $G$ as:
\bea
\label{combgrowth}
G &=& \frac{dR}{dt} \nonumber \\
  &=& D_{out} \frac{C(B(t)) - C_s}{C_{solid} - C_s} K \nonumber \\
  &=& D_{out} \frac{C(B(t)) - C_s}{C_{solid}} K \qquad \mbox{for} \qquad C_s \ll C_{solid}.
\eea
Note that $K$, the rate-constant,
is given by Eqn. (\ref{rateconstant}), expanding  which  we obtain, 
\bea
\label{combgrowth2}
G &=& \frac{dR}{dt} = D_{out}\frac{C(B(t)) - C_s}{C_{solid}} \times \exp \left( \frac{\phi (B(t))}{k_B T} \right) \bigg/\nonumber \\
  && D_{out} B(t)\int_{R(t)}^{B(t)}
\frac{\exp \left( \frac{\phi (r')}{k_B T} \right)}{D_{in}(r') r'} dr'.
\eea
Observing that $\exp[\phi_{B(t)}/k_B T]$ is a constant, we take it to be equal to 1 ($\phi_{B(t)}=0$). We notice
that $D_{out}$ cancels (as it should always do - $D_{in}$ is probably a linear function of $D_{out}$),
and  $D_{in} (r')$ is assumed not to vary within the boundary layer. These lead to:
\bea
\label{combgrowth3}
G = D_{in}\left[ B(t) \int_{R(t)}^{B(t)} \frac{\exp \left( \frac{\phi (r')}{k_B T} \right)}{ r'} dr'\right]^{-1}
   \frac{C(B(t)) - C_s}{C_{solid}}.
\eea

We calculate the above integral using the saddle point approximation, in which, for any function $f(r)$,  
\bea
\label{spdefin}
\int_a^b f(r) dr = f_{max},
\eea
in the leading term where, $f_{max}$ falls in the range between $a$ and $b$, and $f(r)$ dies down both toward $a$ and $b$. This
approximation gets better with $f(r)$ becoming narrower and steeper around $r (f_{max})$ and becomes
exact when $f(r)$ is a delta function at $r (f_{max})$. With this approximation, $r'$ has to be  replaced by
$(R(t)+B(t))/2$ in Eqn. (\ref{barrierpotential}). Consequently, from Eqn. (\ref{combgrowth3}) it follows that
\bea
\label{growthseminal1}
G = D_{in} \frac{C(B(t)) - C_s}{C_{solid}}  \frac{R(t) + B(t)}{2 B(t)} \exp \left( \frac{- A}{k_B T \; \Delta F}\right).
\eea
The point worthy of note here is that we simply considered the numerator of
$e^{\phi (r')/k_B T}/{r'}$ for the saddle-point calculation. This is valid strictly when $A/\Delta F$ is
high compared to $\{B(t) - R(t)\}/R(t)$, i.e, when there is not much variation of $r'$ with respect to the variation of the
exponential function in the range between $R(t)$ and $B(t)$.

Although the above treatment is analytically sound, we still do not know the value of $C_s$, the concentration
at the interface $R(t)$. The calculation is straightforward for temperatures low enough not
to allow desorption at the growth front. In that case, $C_s = 0$, as the monomers in the
polymer chain trying to attach are immobilized as soon as they come in contact with the growth front. For higher temperatures
(i.e., for small super-cooling), desorption is significant, and as a consequence $C_s \neq 0$.
It can be shown that the flux at finite temperatures can be written as,
\bea
\label{fluxfinite}
Fl_{finite} = Fl_{zero} \left( 1 - \exp \left( - \frac{v \Delta F}{k_B T} \right)  \right),
\eea
where, $Fl_{finite}$ and $Fl_{zero}$ are, respectively, fluxes for a finite and zero temperatures
(see, Appendix.I), and $v$ is the volume unit that solidifies. Using the above equation
and identifying $Fl_{zero}$ as a system-specific quantity $\beta$ that can be assigned
a value later (Appendix.I), we immediately reach a different version for the growth law,
\begin{widetext}
\bea
\label{growthseminal2}
G = \beta D_{in} \frac{C(B(t))}{C_{solid}}  \frac{R(t) + B(t)}{2 B(t)} \exp \left( \frac{- A}{k_B T \; \Delta F}\right)
\left[ 1 - \exp \left( - \frac{v \Delta F}{k_B T} \right) \right].
\eea
\end{widetext}
$\beta$ is a diffusion related quantity very weakly dependent on temperature.
The above expression can be presented in terms of the enthalpy  and the super-cooling using the relation,
\bea
\label{freenthalpy}
\Delta F = \Delta H \; \Delta T/ T_m, \qquad \mbox{for small} \; \Delta T
\eea
where, $T_m$ is the equilibrium melting temperature and $\Delta H$ is enthalpy per unit volume of crystallization. Doing so we obtain,
\begin{widetext}
\bea
\label{growthseminalfinal}
G = \beta D_{in} \frac{C(B(t))}{C_{solid}}  \frac{R(t) + B(t)}{2 B(t)} \exp \left( \frac{- A T_m}{k_B T \; \Delta H \; \Delta T}\right)
\left[ 1 - \exp \left( - \frac{v \Delta H \; \Delta T}{k_B T T_m} \right) \right].
\eea
\end{widetext}
This is the most important result in our analysis. We have taken the liberty of applying Eqn. (\ref{freenthalpy}) 
for finite molecular weights also, although it is almost unquestionably valid in the infinite
molecular weight limit only. The volume element $v$ is substantial for large molecules because
it depends on the smallest attaching element to which the attachment-detachment rate-constants 
can be assigned. With a polymer chain, it is an involved analysis\cite{armgol-92} to determine
whether the incorporation of a stem is a one-step process or not.  In our theory, we have 
treated $v$ as a parameter that is believed to be of a value endorsed by experiments. 
We note that the above expression for the growth rate $G$ does not
involve any unknown variable from inside the boundary layer region. Further, it incorporates the factor of
the detailed balance which is present regardless of the size of the molecule. 
The $\exp \left( \frac{- A'}{ T \; \Delta T}\right)$
factor, in addition to the detailed balance factor, has most abundantly been observed in all sorts of nucleation dominated growth
phenomena. As shown above, this factor is recovered in our analysis by using the boundary layer approach 
in which it is embedded in the rate-constant $K$.

In the special case of a dilute solution, the concentration at the boundary layer edge, $C(B(t))$,
is proportional to the initial bulk concentration $C_0$, and so is the concentration inside the
layer. Physically, both the width of the boundary layer ($B(t) - R(t)$) and the diffusion constant inside
the layer ($D_{in}$) will depend on the concentration of the solution.  For higher concentrations,
the polymer molecules will entangle to a higher degree near the growth interface. As a result, the range of the entropic-barrier
which originates from this entanglement would be larger making the boundary layer thicker. The self-diffusivity inside
the boundary layer will also decrease with increasing $C$.  
We assume the conventional theory\cite{doi} of power-law dependence of the self-diffusivity $D_{in}$ on concentration of the form
\bea
\label{dincpower}
D_{in} \sim \frac{1}{C^\alpha},
\eea 
where $\alpha$ is a positive number, to be valid inside the layer. 
Using this relation and the argument above, the growth rate in Eqn. (\ref{growthseminalfinal}) can be written as,
\bea
\label{gpowerc}
G \sim C^{1 - \alpha} \equiv C^\gamma,
\eea
where, $\gamma$ is the concentration exponent. Note that we have left the barrier height $E_h$ as well as the parameter $A$
in the expression of growth rate (Eqn. (\ref{growthseminalfinal}))
as independent of concentration.

\subsection*{B. Numerical results:}

In the previous subsection we discussed the aspects of a continuum theory describing 
the reaction-limited regime of polymer crystallization and analytically derived the
growth laws for a general case of solutions as well as melts.
In this sub-section, we present
numerical calculations for the specific case of dilute solutions to corroborate the analytical theory. 
In the numerical treatment, we address a much wider range of physical situations. As 
mentioned before, even without considering the
complicated barrier forces, the exact solution of the problem involving a moving boundary is
analytically untractable for this circular cylindrical geometry.
Numerical solution is not only capable of dealing with very 
complicated entropic barriers and higher degrees of diffusion in 
different geometries, but also does allow us to analyze competing 
effects resulting from the variation of a single parameter, e.g., 
the molecular weight. For example, as the molecular
weight increases, at one hand it increases the effective super-cooling 
and hence the growth-rate but on the other hand, it enhances the
barrier-effect that impedes the growth. Numerical calculations
deliver the results in a more compact form in these scenarios in which
no single exponent ($\mu$ in Eqn. (\ref{gcmintro})) exists for the whole range of 
the parameter (molecular weight) investigated.

The essence of our numerics is as follows: we have solved
the diffusion equation (Eqn. (\ref{diffeq})) in the region enclosed by the
boundary layer at $B(t)$, the fold surfaces and the walls of the container(Fig. \ref{paper1-fig2}),
subject to the reflective boundary conditions (Eqns. (\ref{bc1}), (\ref{bc2}) )
and the modified mass-balance condition (Eqn. (\ref{blbc}) ). 
The rate constant $K$ (Eqn. (\ref{rateconstant}) ) is determined at each step as a function of the diffusion constant, the
range of the barrier and  the form of the free energy functional related to the
barrier interaction force. Numerically, we
fix the initial radius $R (t = 0)$ at the beginning of the iteration.
The  rate-constant $K$ is calculated subsequently at $B (t = 0)$. The
flux at the edge of the boundary layer ($B(t)$) at finite temperatures
is calculated from a formula related to the flux at zero temperature 
(or a very low temperature - for details of this method see Appendix.I). 
The fluxes at the edge of the 
boundary layer ($B(t)$) and at the interface ($R(t)$) are equated (or, at the least, made proportional)
Once the concentration ($C$) for the whole space in the
bulk solution is updated,
the new radius is calculated from the old radius using the mass balance
equation (Eqn. (\ref{massbal})). Note that a rigorous integration in
Eqn. (\ref{rateconstant}), instead of a saddle-point approximation, is
possible in the numerical scheme.  

There is an important aspect in the simulation that warrants a note.
The growth of a single crystal in a dilute solution
involves a moving boundary in a time-dependent diffusive environment.
These problems are generally treated as moving boundary value problems
in the literature. Till date, there is no analytical solution available
for this problem in a finite circular cylindrical system with arbritrary 
concentrations as the diffusion variable. In our numerical program, we have employed a technique
in which the position-grid in the radial direction ($r$ values) has
been adjusted at each step so that one grid-point always coincides with 
the edge of the boundary-layer, at which location the rate constant boundary condition
(Eqn. (\ref{blbc})) is enforced.

As mentioned before, three major factors affecting the growth rate and typically discussed in the literature
have been investigated in our numerical work on dilute solutions. They are the crystallization temperature $T_c$ (in terms of the 
under-cooling or super-cooling $\Delta T$) of the system, concentration ($C$) of the bulk
solution and the 
moleculer weight ($M_w$) of the crystallizing polymer. Polymer crystallization
typically being a very slow process no temperature gradient is assumed to be present 
in the solution.

\subsubsection*{1. Under-cooling, $\Delta T$:}

Linear homopolymers in a dilute solution are generally crystallized by 
reducing the temperature below the equilibrium melting temperature or,
in other words, by undercooling the solution. Experiments throughout have
shown that  the growth rate depends heavily on the degree of under-cooling,
and it has been orders of magnitude lower than what is expected in a
diffusion limited growth. At the same time, it has been observed that
the rate changes by orders of magnitude for a relatively small change
in temperature. All these evidences strongly suggest that polymer
crystal growth in dilute solutions is a reaction-dominated phenomena,
the reaction at the growth face being critically dependent on the
degree of under-cooling $\Delta T$.

\begin{figure}[ht] \centering
\hspace*{1cm}{\epsfxsize= 12cm \epsfbox{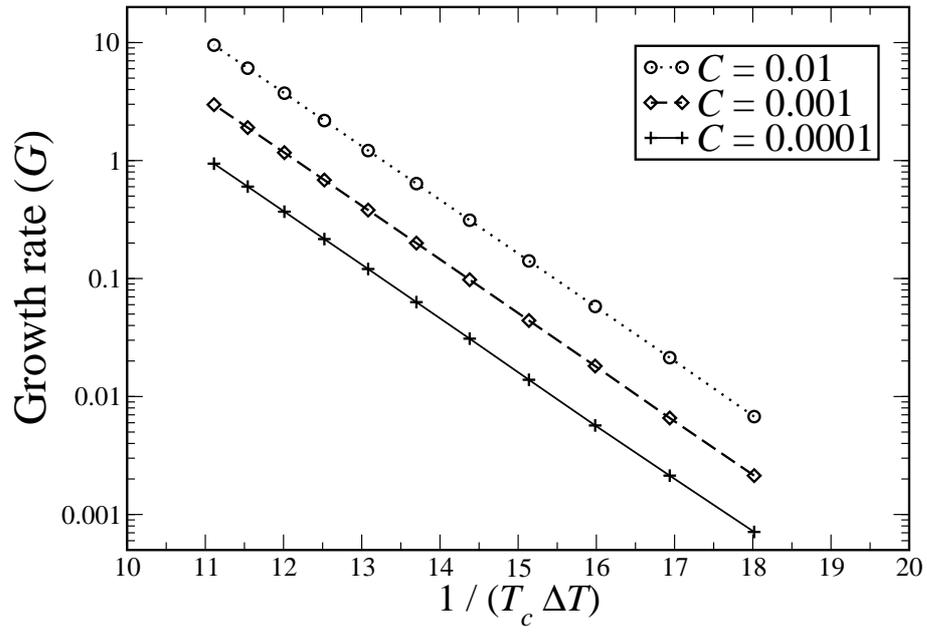}}
\bigskip
\vspace*{-0.5cm}
\vspace*{5cm}
\caption{Kundagrami \it{et al.}, JCP}
\label{paper1-fig5}
\end{figure}

The saddle-point analysis in the last section had shown that the growth
rate can be written as 
\bea
\label{saddle-result}
G \sim C_0^\gamma \exp^{ - \left(A'/T_c \Delta T \right)} \left[ 1 - \exp^{ - \left(B' \Delta T \right)} \right] ,
\eea 
where, $T_c$ is the crystallization temperature, $A'$ and $B'$ are system parameters not too critically dependent on 
temperature, concentration or molecular weight,  
$C_0$ is the initial concentration, $\gamma$ is the concentration exponent and $\Delta T = T_m - T_c$, where,
$T_m$ is the equilibrium melting temperature of that specific polymer.
We did an explicit numerical calculation of the growth rate
for three different concentrations, $C = 0.01,\; 0.001$ and  $0.0001$, all of which fall in the dilute
regime. The growth rate $G$ is plotted against the factor $1/T_c \Delta T$
in Fig. \ref{paper1-fig5}. The temperature range considered 
was from 15 degree to 25 degree below $T_m$. Comparing this result with
experiments by Keller and Pedemonte\cite{kelped-73}, we
infer that our theory has excellent compliance with the generic experimental 
results as the graphs are  good straight lines with the growth rate
increasing by at least three orders of magnitude for an increase in
undercooling from 15 to 25 degrees. As expected, the slope of the 
straight lines remain unaltered for all concentrations implying
no dependence of concentration on the prefactors $A'$ and possibly $B'$. The 
numerical result corroborates the well-known experimental observation that
the nucleation factor entirely suppresses the detailed balance factor 
(Eqn. (\ref{saddle-result})) for the given range of moderate under-coolings.

\begin{figure}[ht]  \centering
\begin{minipage}{15cm}
\vspace*{1.2cm}
\hspace*{0.0cm}\textsf{\textbf{(a)}}\\
\vspace*{-1.0cm}
\hspace*{1.5cm}{\epsfxsize= 10cm \epsfbox{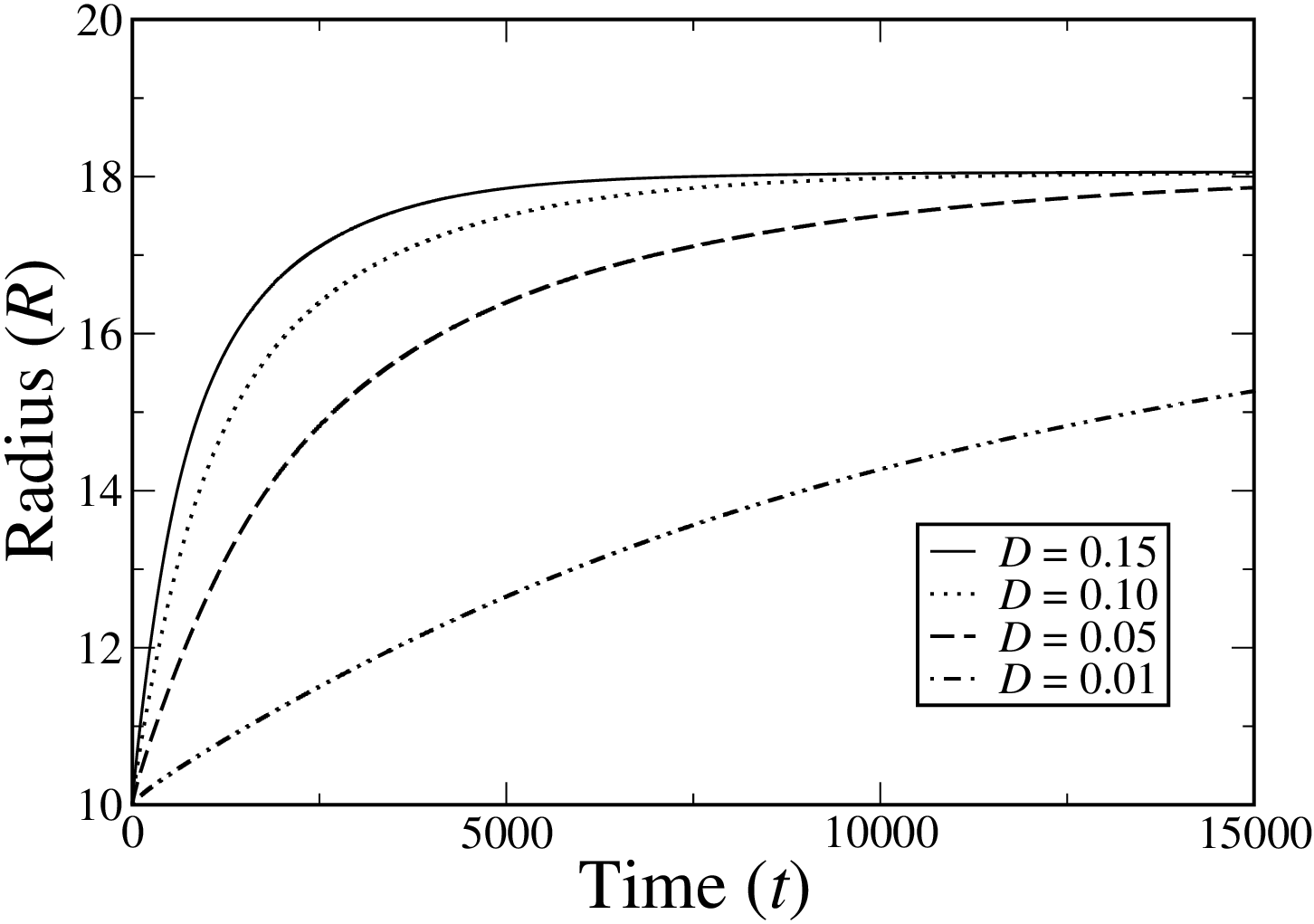}}\\
\vspace*{1.2cm}
\hspace*{0.0cm}\textsf{\textbf{(b)}}\\
\vspace*{-1.0cm}
\hspace*{1.5cm}{\epsfxsize= 10cm \epsfbox{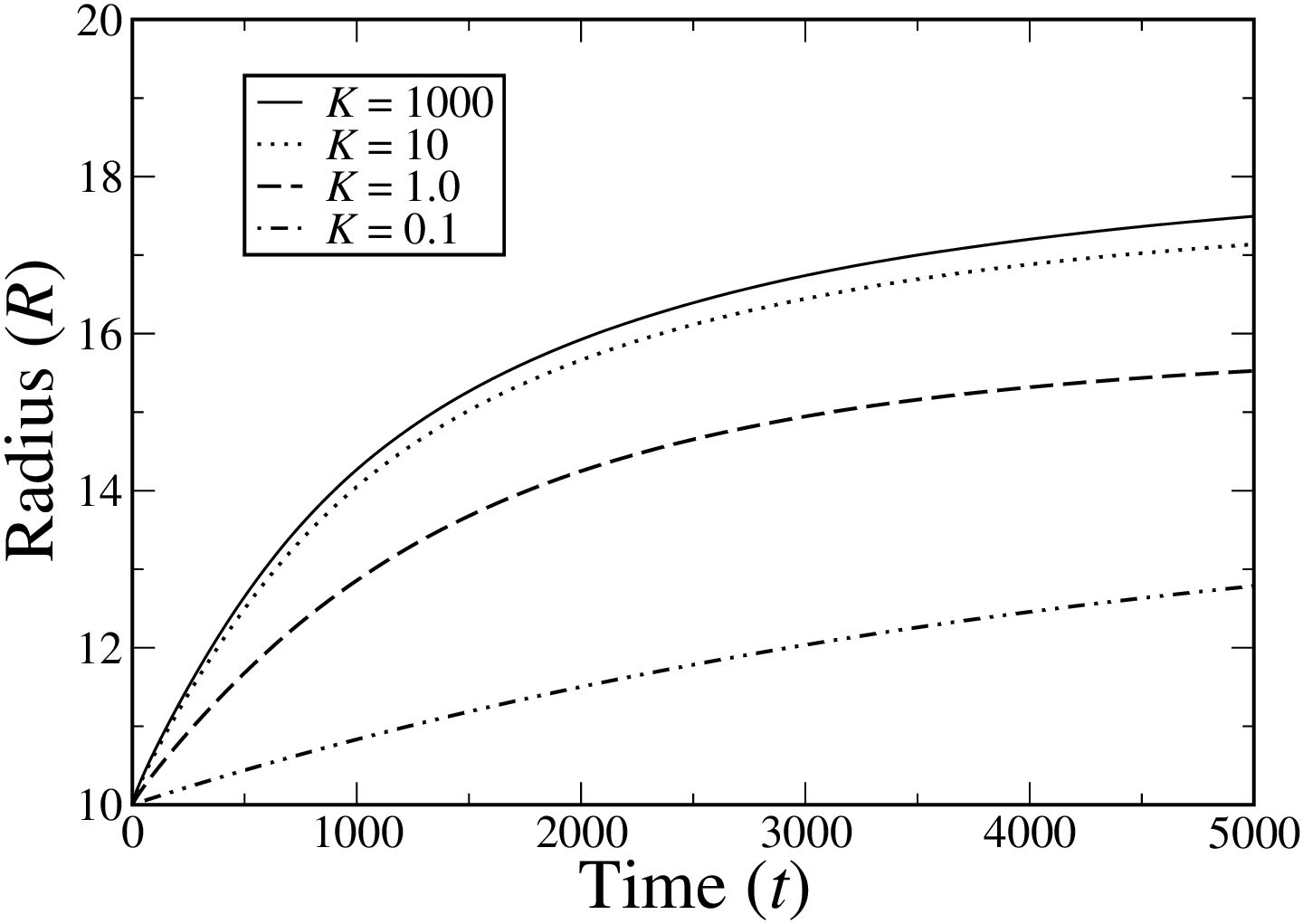}}\\
\vspace*{1.2cm}
\hspace*{0.0cm}\textsf{\textbf{(c)}}\\
\vspace*{-1.0cm}
\hspace*{1.5cm}{\epsfxsize= 10cm \epsfbox{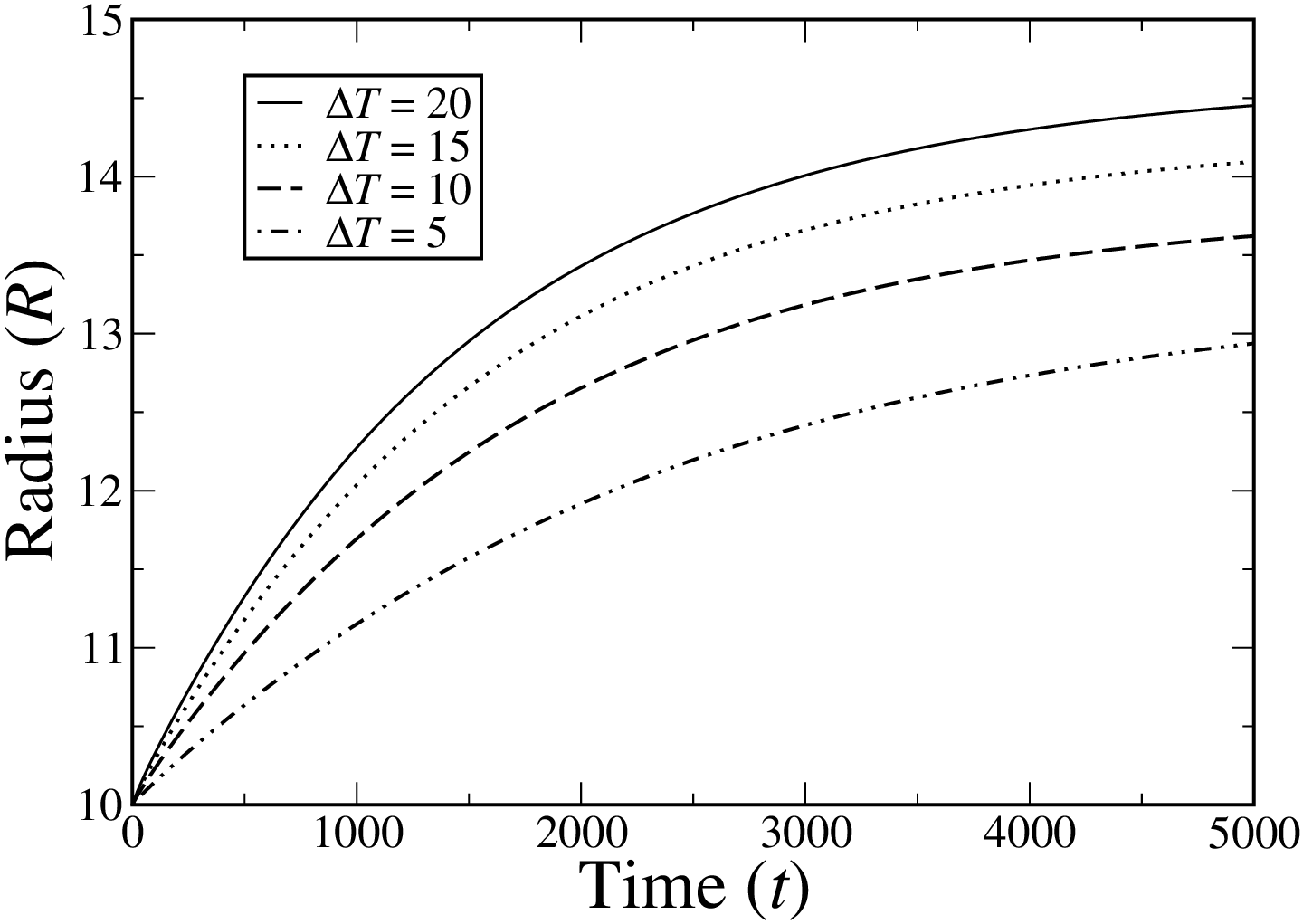}}\\
\caption{Kundagrami \it{et al.}, JCP}
\label{paper1-fig6}
\end{minipage}
\end{figure}

To compare the above mentioned growth governed by nucleation and entropic
barrier to one that is not (and hence allows
uninhibited diffusion-limited attachment), we calculated crystal growth for 
very large values of both the rate-constant $K$ (implying effectively zero barrier) and
under-cooling $\Delta T$ (implying no dissolution). The rate-constant in these
calculations is fixed externally and does not depend on the under-cooling. The growth rate is plotted 
for four different diffusivities in Fig. \ref{paper1-fig6}a. 
A sufficiently long  time-range of the growth inside the container
has been captured, and hence it shows the effect of depletion of available
material during the later part of the growth. The initial growth rate, free from the effect
of the finite boundary, is found to be orders of magnitude higher than the
nucleation dominated rates in Fig. \ref{paper1-fig5}. To
show the effect of the barrier, we have performed similar growth calculations
for lower values of $K$ and plotted the growth ($R$) with time ($t$) in 
Fig. \ref{paper1-fig6}b. We notice that the growth rate
saturates to the value corresponding to diffusion-control for high values
of $K$. We further calculated the growth for small under-coolings ($\Delta T$ = 5 to 20 degrees)
and plotted it ($R$) with time ($t$) in Fig. \ref{paper1-fig6}c.
It is evident from the plot that the effect of desorption, which is a significant fraction of adsorption
at moderate under-coolings, slows down the process. But, as we notice from the slopes of the four
curves in Fig. \ref{paper1-fig6}c, the growth rate changes by
only a factor of two for a 10 degree change in under-cooling. This is way less than the factor
of $10^3$ present in the nucleation dominated growth in Fig. \ref{paper1-fig5}.

\subsubsection*{2. Concentration, $C$:}

In case of diffusion limited growth for small molecules in a solution, the growth rate is
generally proportional to concentration. The fact that the growth rate
depends on concentration raised to the power some number less than unity implies
the presence at the growth front of a barrier, the strength of which depends 
on the concentration itself. In mathematical terms, $\gamma$ in
\bea
\label{gamma}
G \sim C^\gamma
\eea
has been observed to be less than one in most cases\cite{kelped-73,cooman-75}.

As per the analytical discussion in section III.A, Eqn. (\ref{massbal}) and
Eqn. (\ref{blbc}) can be simplified to show that the growth rate follows
\bea
\label{gkc1}
G &\sim& \mbox{flux} \nonumber \\
&\sim&  D_{out} \frac{\partial C}{\partial r} \nonumber \\
&\sim& D_{out} K C, 
\eea
where we have left out the temperature factors for the time-being (see, Eqn. (\ref{saddle-result})).
For generic linear homopolymers in solution, the dissolution temperature $T_d$
does not vary much with concentration, especially in the higher molecular weight limit 
(a specific example of polyethylne in xylene is given in \cite{paskho-84}).
Therefore, the equation above implies that unless the rate-constant $K$ is a function of 
$C$, the growth rate is simply proportional to it and $\gamma$ is unity. For small molecules, $K$ is 
unaffected by the concentration of the solution, but for molecules as large
as polymers, we suggest that the concentration has an effect on the degree of crowding of
molecules at the growth front, and hence it affects the value of $K$. A simplified 
version of Eqn. (\ref{rateconstant}) sheds more light on the effect
of various quantities, especially concentration,  on the growth rate (section III.A).
Considering the concentration dependence of the self-diffusivity inside the boundary layer region, a simple
dependence of growth rate on concentration has already been obtained (Eqns. (\ref{dincpower}) and (\ref{gpowerc}) ) in this article. 
In the  dilute solution limit -  for example for polystyrene polymer in benzene with the initial concentration
being less than $0.01$ -  $\alpha$ is close to zero and the self-diffusivity $D_{in}$ does not vary much with
the density of monomers\cite{herleg-79}. On the other hand, if the concentration is close to $0.1$ or higher,
the Rouse regime sets in and $\alpha$ is close to $2$. In our model, we expect the concentration
inside the layer to be somewhere in this range, and hence the value of $\alpha$ to be between $0$ and
$2$, with the most probable values being around $0.5$ to $1$. For example, if we take the value
of $\alpha$ to be $0.5$, $\gamma$ is $0.5$ which is an acceptable
result supported by experiments\cite{kelped-73,cooman-75}. For  many experiments in higher concentrations,
$\gamma$ has been found to be very low (0.1 to 0.2). 
However, if we notice that $\gamma = 1 - \alpha$ in Eqn. (\ref{gpowerc}),
it is apparent that the value of $\alpha$, depending on the concentration of
the solution would affect the growth rate in a very simple but significant way.
Assumption that the concentration inside the barrier layer renders
$\alpha$ to be  close to but little less than unity immediately
results in a growth rate highly insensitive to concentration.
The data in Ref.\cite{herleg-79}
is very much supportive of the  above hypothesis.
To obtain a wider range for the value of $\gamma$ as observed in experiments, one has to
consider other factors as mentioned above, which, at this point, are beyond the scope of
this work. 

\begin{figure}[ht] \centering
\hspace*{1cm}{\epsfxsize= 12cm \epsfbox{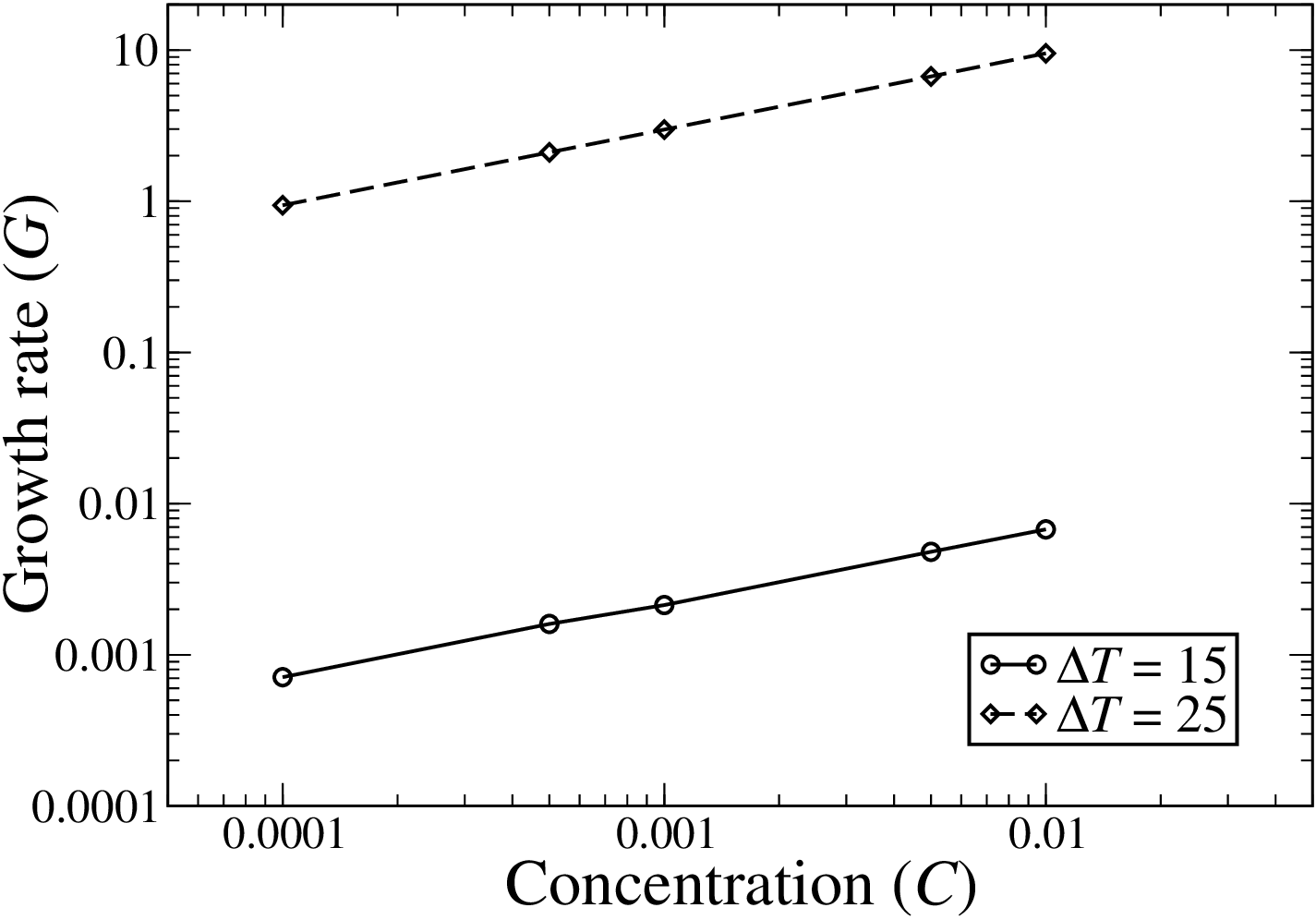}}
\bigskip
\vspace*{-0.5cm}
\vspace*{5cm}
\caption{Kundagrami \it{et al.}, JCP}
\label{paper1-fig7}
\end{figure}

The growth rate as a function of concentration is plotted for two 
under-coolings in Fig. \ref{paper1-fig7}. The higher
growth-rate (dashed) line is for higher under-cooling ($\Delta T = 25$) and
the lower  (solid) line is for lower under-cooling ($\Delta T = 15$). 
The value of $\alpha$ used in Eqn. (\ref{gpowerc}) is 0.5. No
dependence of growth rate on the crystallization temperature ($T_c$)
has been observed. The value of the concentration exponent $\gamma$
is found to be $0.5$ in both cases.

\subsubsection*{3. Molecular weight, $M_w$:}

Unlike temperature and concentration, the molecular weight affects the
growth rate non-monotonically as it is almost
impossible to find one single molecular weight exponent for 
the whole range of it. Assuming all other things remain the same, the growth
rate depends on concentration with a fixed exponent. This is applicable for
a good range of concentration large and small molecule systems alike
because of the diffusion control which is proportional to $C$. A change in 
molecular weight for flexible macromolecules, however, changes their equilibrium melting 
temperature ($T_m$), and hence changes the effective super-cooling 
($\Delta T = T_m - T_c$) when experiments are performed on a isotherm ($T_c$).
This change in $\Delta T$ is not linear with $M_w$ and, therefore,
is the specific reason behind the non-monotonic molecular weight dependence (logarithmic) 
of the growth rate. In light of our theory, the other major effect 
we propose is that increasing molecular weight decreases the self-diffusion constant $D_{in}$
inside the boundary layer slowing down the growth rate. These two competing
effects render the molecular weight dependence of the growth rate to be complicated and
analytically complex. 
In a typical system, for example for polyethylene
single crystals in xylene solution, the growth rate increases with the molecular weight for small
$M_w$'s and later hits a plateau or even decreases for higher $M_w$'s depending on the concentration
of the solution. 

To numerically calculate the growth rates as a function of the size of the molecule, we use
the well-known empirical expression for the equilibrium melting temperature as a function of the molecular weight.
The melting or dissolution
temperature of a finite molecular weight polymer can be written as\cite{wun-80},
\bea
\label{meltmol}
T_m = T_m^0 - \frac{E}{M_w},
\eea  
where,  $T_m^0$ is the equilibrium melting temperature for infinite molecular weight and  $E$ is a constant 
for the polymers of the same series with different molar mass.
The above formula is a Gibbs-Thomson type expression relating the melting temperature for
finite and infinite systems but is still phenomenological, and input for $E$ must be obtained from real systems.
$T_m^0$ and $E$ are available for several systems in the literature\cite{wun-80,paskho-84}. To account
for the other effect, in accordance with the conventional polymer theories, the following dependence of the self-diffusion constant $D_{in}$
on the molecular weight can be assumed:
\bea
\label{dinmol}
D_{in} \sim \frac{1}{{M_w}^x},
\eea
where, $x$ is unity in the Rouse regime. We can predict the trend of the growth rate versus molecular weight
curve qualitatively from above two relations. For low molecular weights, a change in $T_m$ (hence $\Delta T$, 
the effective super-cooling) with $M_w$ will be significant implying large change in the growth rate due to the
$\exp [-1 / T_c \Delta T]$ factor. The growth rate would, therefore, increase rapidly with $M_w$ in the lower
range. For larger molecular weights, $T_m$ will saturate to the value of $T_m^0$.
If $M_w$ is further increased, Eqn. (\ref{meltmol}) will cease
to affect the rate and Eqn. (\ref{dinmol}) will take over. The growth mechanism, therefore, will be progressively retarded. 
The curve might have a plateau depending on the constants involved in the above two equations. 

\begin{figure}[ht] \centering
\hspace*{1cm}{\epsfxsize= 12cm \epsfbox{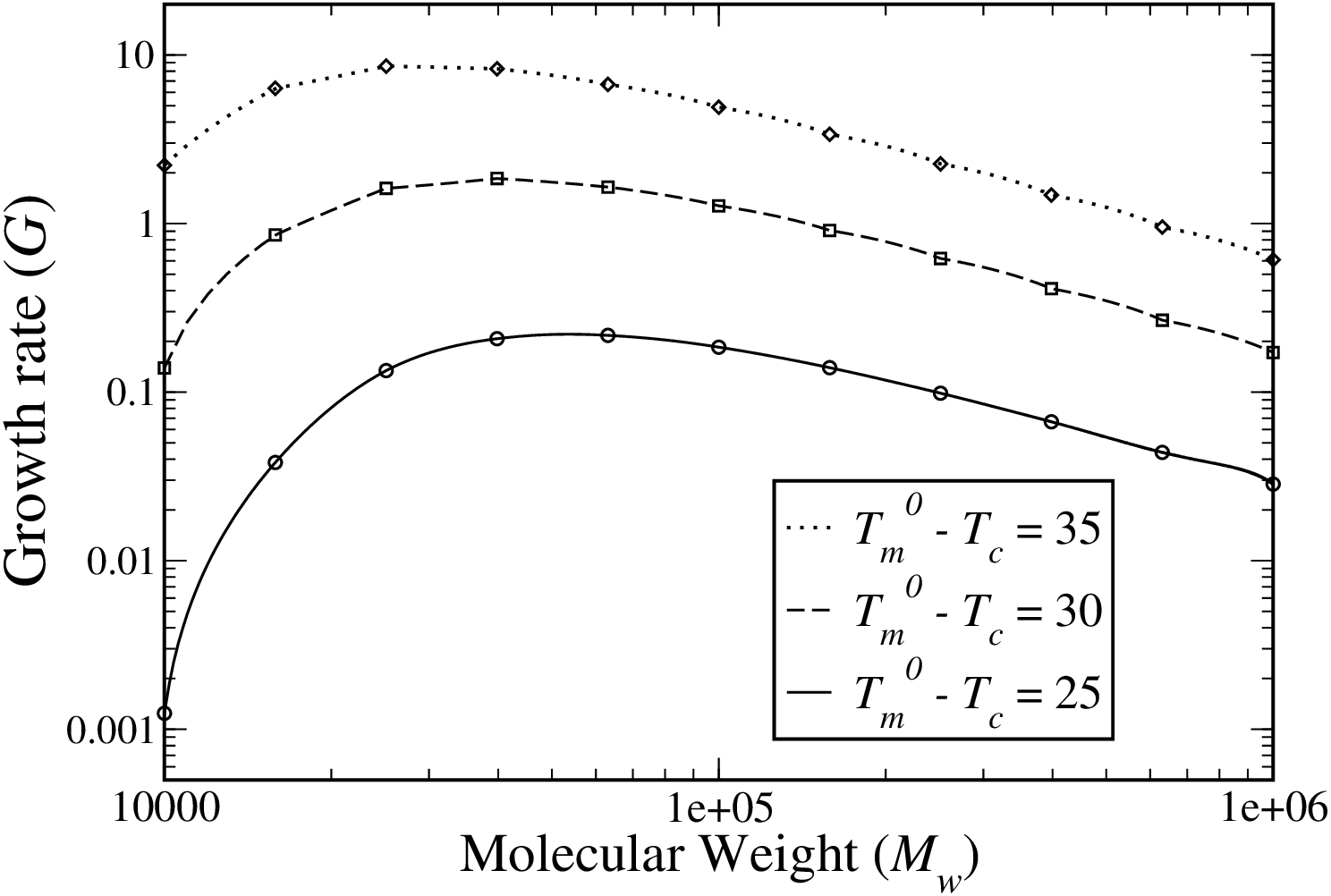}}
\bigskip
\vspace*{-0.5cm}
\vspace*{5cm}
\caption{Kundagrami \it{et al.}, JCP}
\label{paper1-fig8}
\end{figure}

Using the relationships mentioned above in our numerics, we have plotted the growth rate ($G$) against the molecular weight ($M_w$)
for three crystallization temperatures ($T_c = T_m^0 - 25, T_m^0 - 30, T_m^0 - 35$) 
(Fig. \ref{paper1-fig8}).  The rate-curve isotherms agree 
reasonably well with the qualitative argument presented above and with experiments\cite{cooman-75}. 
We have chosen the concentration $C$ to be $0.001$,
which is at the middle of the range we considered in this work. The shapes of the isotherms do not change
significantly with concentration, although the absolute values of the growth rates do. There are no
plateau in these particular curves but various forms and shapes of these isotherms
are obtained in our numerics by changing the constants, especially $E$, as mentioned above.

The effect of the molar mass of the polymers on the concentration exponent $\gamma$ has been
investigated extensively\cite{cooman-75} in the literature and is worthy of analysis. For a substantial range of crystallization temperatures,
$\gamma$ goes down with $M_w$ for a fixed crystallization temperature, $T_c$. In other words, the effect that renders the
growth rate more insensitive to concentration increases with the molecular weight of the polymer. 
Within the purview of our theory, this implies of two possibilities: a) the self diffusion constant
$D_{in}$ inside the barrier layer still follows Eqn. (\ref{dincpower}) but $\alpha$ increases
with $M_w$ and b) the range of the barrier layer ($R(t) < r <B(t)$), and therefore, the limit
of integration in the expression of the rate constant $K$ (Eqn. (\ref{rateconstant})), 
has a concentration dependence of the form $\sim {C_0}^y$, where, $y$ 
increases with $M_w$. Either of the effects or both can be present. Once this hypothesis
is proved to be valid, one might reproduce the growth rate versus molecular weight
curves at different isotherms for different concentrations\cite{cooman-75} and
show that growth becomes more inhibited for higher concentrations for the same
range of molecular weight. At this point, this analysis is in a speculative level and we refrain from making
a conclusive remark on this.

\section*{IV. CONCLUSIONS}

\setcounter{section}{4}
\setcounter{equation}{0}

The key conclusion reached from our theoretical formalism is that the flexibility or conformational entropy
of the polymer chains is the distinctive rate-controlling factor that separates polymer crystallization 
from small-molecule crystallization. There is a significant entropic contribution to the free energy ($E-TS$)
of the ordering process due to the reduction of available conformations of the polymer chains. The barrier
layer theory we propose addresses the entropy factor in addition to the energy factor, which is the only
factor the LH theory and its modifications consider. This formalism recovers both the nucleation dominated
limit for large flexible molecules (Eqn. (\ref{gdtintro})) and diffusion controlled limit for small molecules 
(Eqn. (\ref{smallmolintro})) in one single expression. In addition, unlike in the LH theory, this model
considers one fixed thickness and still recovers the '$\exp(-1/T \Delta T)$' factor most abundantly observed in 
polymer crystallization literature.    

The free energy barrier related to the loss of entropy is assumed to be prevalent in a narrow layer
adjacent to the growth front, referred to as  as the barrier layer or the growth zone. The temporal congestion
of entangled chains occurs only in this region creating an impedance    
to the lateral growth process. 
The conclusive analytical result in our theory is summarized in the following expression:
\bea
\label{growthconclusion}
G &\sim& C_0 D_{in} \exp\left( -\mathcal{P}/ T_c \Delta T \right) \nonumber \\
&&\left[ 1 - \exp\left( \Delta H \Delta T / k_B T_m T_c\right) \right], 
\eea      
where, $G$ is the linear growth rate, $C_0$ is the initial concentration, $D_{in}$
is the diffusion coefficient inside the boundary layer, $\Delta H$ is the enthalpy of fusion,
$\Delta T$ is the under-cooling, $T_m$ is the melting temperature at the finite molecular weight, 
$T_c$ is the crsytallization temperature and $\mathcal{P}$ is a parameter which depends on
the details of the entropic barrier as well as on $\Delta H$, $T_m$ and $T_c$. The
above equation captures both the barrier control ($G \sim \exp (- 1/\Delta T)$) and diffusion
control ($G \sim \Delta T$, for small $\Delta T$). Both $D_{in}$ and the '$\exp (- 1/T_c \Delta T)$' term become
insignificant when the barrier layer thickness is negligible, i. e., for low molecular weight
polymers in solutions and small molecules. 

Physical variables such as the diffusivity and its dependency on concentration and molecular weight
have been assumed to follow a different dynamics in the dense growth zone. Following this, the above
growth law predicted growth rates to be weakly or marginally dependent on concentration (the exponent
$\gamma$ in $G \sim C^\gamma$ is $<$ or $\ll 1$, respectively). These agree well with the
experimental results which at the very first place suggested a barrier control near the growth front.
In addition, considering a phenomenological
relation between the melting temperature and molecular weight, we recovered typical experimental
results for the dependency of growth rates on molar mass.

Our formalism is qualitatively different from the classical formulations of polymer crystallizations
based only on energy arguments.  Although the flexible nature of long
chain molecules gives rise to various complex and diverse growth forms,
we have shown that the growth can be modelled by simple physical processes considering a few typical features 
of crystallization. However, more detailed work is needed to determine the true nature
of the entropic barrier, and to make quantitative comparison with experiments.

A few observations related to our work are noteworthy. First, we have retained the concept of a surface free energy
for the crystal-liquid interface assuming the basic arguments
for the origination of the energy to be still valid. The degeneracy due
to many possible and energetically equivalent options for the stems to arrange themselves on
the surface gives rise to the surface free energy at the very first place.
In our model, the stems are still subject to this degeneracy $at$ the front after they negotiate
the entropic barrier layer $adjacent$ to the front.

Second, we turn our attention to the important parameter $\mathcal{P}$ in Eqn. (\ref{growthconclusion}). This parameter has a direct
correlation to the minimum volume unit $v$ that solidifies in 'near equilibrium' (the detailed balance factor).
Therefore, $A$ contains the thickness, $l$, of the growing lamella. It might be possible to formulate the
effect of quench depth $\Delta T$ on the thickness, broadly in line with the traditional theories. Our
theory, however, has shown that a thickness dependent growth rate is not a necessary prerequisite to generate the 'nucleation' factor
(see also Appendix.II). 

Third, for a fixed molecular weight $M_w$, the concentration exponent 
$\gamma$ has been observed in experiments to increase slightly with $T_c$, the crystallization temperature. We did not address
this behaviour in our theory.

\noindent
\section*{ACKNOWLEDGMENT}

Financial support for this work was provided by the National Science Foundation Grant DMR-0605833 and
MRSEC at the University of Massachusetts, Amherst.
 

\section*{APPENDIX.I}

\setcounter{section}{5}
\setcounter{equation}{0}

In this section we present a simple derivation for the expression of particle flux at the growth 
front at finite temperatures (as described in Sec.III). As desorption is important at a finite temperature 
situation, we consider the inward and the outward fluxes separately and the difference is the
net flux of particles that finally attach to the growing surface. Representing the inward and 
outward fluxes as $Fl_{in}$ and $Fl_{out}$, respectively, and assuming the 
detailed balance to be valid at the interface, we write,
\bea
\label{detbal}
\frac{Fl_{out}}{Fl_{in}} = \exp \left( - \frac{v \Delta F}{k_B T} \right),
\eea
where, $\Delta F$ is the gain in free energy per unit volume after solidification and $v$ is
the volume unit that solidifies. $\Delta F$
is, in general, proportional to super-cooling $\Delta T$ for small $\Delta T$'s and diverges with $\Delta T$.
Assuming the fluxes $Fl_{in}$ and $Fl_{out}$ to be $\beta e^{-F_1/K_B T}$ and $\beta e^{-F_2/K_B T}$,
respectively, where, $F_1$ and $F_2$ are the activation energies for attachment and detachment, respectively,
and assuming $\beta$ to be an arbitrary, temperature independent quantity,  
we can express the net flux, $Fl_{total} = Fl_{in} - Fl_{out}$, at the interface as
\bea
\label{netfluxboundary}
Fl_{total} \equiv \frac{\partial C}{\partial r}\bigg|_{r = R(t)} 
= \beta \left( 1 - \exp \left( - \frac{v \Delta F}{k_B T} \right)  \right).
\eea
For $\Delta T \rightarrow \infty$ (at very low temperatures) the rate of desorption
at the interface is negligible to the rate of absorption and almost all the molecules
that attach to the interface stick there permanently to be a part of the full-grown
crystal. In this limit, the perfect sink boundary condition ($C = 0$) applies at the 
interface because all the diffusing molecules are immobilized as soon as they come
in contact with it. This implies that
\bea
\label{beta}
\beta = \frac{\partial C}{\partial r}\bigg|_{r = R(t), C_s = 0}.
\eea

The factor $e^{-F_1/K_B T}$ in $Fl_{in}$ is very weakly dependent on temperature can be
assumed to be a constant for a moderate range of temperature. Considering that and 
combining Eqns. (\ref{netfluxboundary}) and (\ref{beta}) we reach a general 
expression for the flux at the interface at finite temperatures,
\bea
\label{netfluxfinal}
&&\frac{\partial C}{\partial r}\bigg|_{r = R(t), \mbox{general}} \\ \nonumber 
&=&\frac{\partial C}{\partial r}\bigg|_{r = R(t), C_s = 0} \left( 1 - \exp \left( - \frac{v \Delta F}{k_B T} \right)  \right),
\eea
which is applied in the algorithm described in Sec.III.

\section*{APPENDIX.II}

\setcounter{section}{6}
\setcounter{equation}{0}

In this section we derive the growth rates for different regimes using a fundamental 
surface nucleation theory and compare the results with that of the Lauritzen-Hoffman
theory. The theory presented here is much simpler and does not require to conceive a 
cut-off minimum and a distribution of thickness as in the LH theory. We start
with the basic expression for the free-energy change for solidification of $m$ number
of stems of width (the lateral direction of 'substrate completion') $a$, thickness 
(the growth direction normal to the interface) $b$ and length (lamellar thickness) $l$.
The free energy consists of the conventional bulk gain and surface loss terms and we have 
\bea
\label{deltagalt}
\Delta G = - m a b l \Delta F + 2 b l \sigma + 2 m a b \sigma_e,   
\eea 
where, $\sigma$ and $\sigma_e$ are the lateral and fold surface free energies, 
respectively. To determine the critical free energy $(\Delta G)^\star$, the
expression in Eqn. (\ref{deltagalt}) is maximized with respect
to both variables  in the system - the number of stems in a nucleus $m$ and 
the thickness of the lamella $l$. Consequently, $\partial \Delta G/\partial m = 0$
and $\partial \Delta G/\partial l = 0$ imply, respectively, that
\bea
\label{deltagmax}
l^\star &=& \frac{2 \sigma_e}{\Delta F} \qquad \mbox{and} \nonumber \\ 
m^\star &=& \frac{2 \sigma}{a \Delta F}, 
\eea
where, $l^\star$ and $m^\star$ denote the critical values (at $(\Delta G)^\star$) 
for the quantities. Substituting the critical values in the expression of 
$\Delta G$, we obtain
\bea
\label{deltagstar}
(\Delta G)^\star = \frac{4 \sigma \sigma_e b }{\Delta F},
\eea
which has been the essential conclusion for the LH theory. To illustrate the comparison
we recall that for the small molecule nucleation theory, the growth rate $G_r$ is
a function of the nucleation rate $i$, which in turn depends on the height of
the nucleation barrier ($(\Delta G)^\star$ in our case) in the following way,
\bea
\label{nucrate}
i \sim \exp \left( - \frac{(\Delta G)^\star}{k_B T} \right).
\eea
Subsitution of the expression for $(\Delta G)^\star$ (Eqn.\ref{deltagstar}) in the
above equation yields,
\bea
\label{nucratelhalt}
i \sim \exp \left(  - \frac{4 \sigma \sigma_e b}{\Delta F\:\: k_B T} \right),
\eea
which is the more well-known LH result. The regime I and II expressions can
be obtained following the arguments in line with the standard nucleation
theories. For the two-dimensional geometry of polymer crystal growth, in regime I,
the growth rate $G_r \sim i$ and in regime II, $G_r \sim i^{1/2}$. In
conjunction with the expression for the nucleation rate $i$ and assuming that the
analysis is tenable to large molecules, the growth rates immediately
explain the slope change of the $\log G$ vs. $1/T_c \Delta T$ straightline by  a
factor of two when the growth process shifts from regime I to II.

\section*{FIGURE CAPTION}
\pagestyle{empty}
\begin{description}

\item[Fig. 1.:] Concentration ($C$) and free energy ($\phi$) profiles as a functions of the distance from the
interface ($r$). Polymer molecules accumulate (top) in the boundary layer
region, the concentration increases substantially (the graph is not to scale - $C$ is order
of magnitude higher in the layer region) close to the interface (middle) and
a free energy related to the entropic barrier is created (bottom) near the interface.

\item[Fig. 2.:] Schematic of the tabular growth of polymer crystals from dilute solution.
Flat surfaces of the right circular cylider normal to $z$ axis are the fold surfaces with
no growth. The solid-solution interface is at $R(t)$ and the edge of the dense boundary
layer is at $B(t)$. 

\item[Fig. 3.:] Schematic of the chain attachment and the related free energy:
(top) different stages of attachment of one single chain to the growth front.
(bottom) the free energy as a function of the number of monomers attached. 
There is a critical number above which the process is downhill.

\item[Fig. 4.:] A model free energy barrier as a function of the distance from
the interface. The shape is chosen as a parabola in the theory. 

\item[Fig. 5.:] The temperature dependence of growth plotted as growth rate ($G$) vs. $1/(T_c \Delta T)$
for three concentrations: $C = 0.0001$(plus), $C = 0.001$(diamond), $C = 0.0001$(circle).
In the log-scale they are excellent straight lines for all three concentrations. The abscissa
is multiplied by $10^5$ and the ordinate has arbitrary units.

\item[Fig. 6.:] Growth or size (radius $R$) as a function of time $t$
in various conditions: (a) diffusive with no barrier (rate constant $K$ is very high) and
at a very high super-cooling $\Delta T$ (no dissolution) for diffusion co-efficients ($D$):
0.01, 0.05, 0.10, 0.15.
(b) at a very high super-cooling $\Delta T$ (no dissolution) for various barrier strengths:
$K = 0.1, 1, 10, 1000$ with $D = 0.1$. Notice that the growth rate saturates with increasing $K$.
(c) for a fixed barrier strength ($K = 1.0$) and diffusion co-efficient ($D = 0.1$) for
various moderate super-coolings, $\Delta T$ : 5, 10, 15, 20. Dissolution is significant
at these super-coolings. Depletion of material is reflected as the graphs flatten for
long times. Graph (c) indicates that, for cases where $K$ is not dependent on $\Delta T$,
growth rate changes by less than one order for a typical experimental range of $\Delta T$.

\item[Fig. 7.:] The concentration dependence of growth plotted as growth rate ($G$) vs.
concentration ($C$) for two super-coolings: $\Delta T = 15$(circles), $\Delta T = 25$(diamonds).
In the log-scale they are excellent straight lines with slope ($\gamma$) being close to
0.5 for both $\Delta T$. No dependence on the crystallization temperature, $T_c$, is allowed in the theory.

\item[Fig. 8.:] The molecular weight dependence of growth plotted as growth rate ($G$) vs.
 ($M_w$) for three  crystallization temperatures: $T_c = T_m^0 - 25$(circles), $T_c = T_m^0 - 30$(squares),
and $T_c = T_m^0 - 35$(diamonds). $T_m^0$ is the equilibrium melting temperature in the infinite molecular
weight limit. The concentration is chosen to be $C = 0.001$. We notice that there is no fixed
molecular weight exponent for the growth for the entire range of $M_w$. In fact, generally, the
growth rate increases with $M_w$ initially (due to the increase of effective super-cooling) and
decreases later (due to the entropic force near the barrier).

\end{description}

\newpage

\newpage

\newpage

\newpage

\newpage

\newpage

\newpage

\newpage

\end{document}